\documentclass[12pt,draftcls,onecolumn]{IEEEtran}
\usepackage{mathrsfs}
\usepackage[dvips]{graphicx}
\usepackage{color}
\usepackage{subfigure}

\usepackage{amsfonts}
\usepackage{amsmath}
\usepackage{amssymb}
\usepackage{amsthm}
\usepackage{cite}
\usepackage{url}
\usepackage{slashbox}
\usepackage{algorithm}
\usepackage{algorithmic}

\graphicspath{{/}}

\begin{document}

\title{On Cost-Effective Incentive Mechanisms in Microtask Crowdsourcing}

\author{\IEEEauthorblockN{Yang Gao, Yan Chen and K. J. Ray Liu}
\IEEEauthorblockA{\\Department of Electrical and Computer Engineering,\\University of Maryland, College Park,
MD 20742, USA\\ E-mail: \{yanggao, yan, kjrliu\}@umd.edu}}


\maketitle

\newtheorem{proposition}{Proposition}
\newtheorem{definition}{Definition}
\newtheorem{theorem}{Theorem}
\newtheorem{corollary}{Corollary}

\begin{abstract}
While microtask crowdsourcing provides a new way to solve large volumes of small tasks at a much lower price compared with traditional in-house solutions, it suffers from quality problems due to the lack of incentives. On the other hand, providing incentives for microtask crowdsourcing is challenging since verifying the quality of submitted solutions is so expensive that will negate the advantage of microtask crowdsourcing. We study cost-effective incentive mechanisms for microtask crowdsourcing in this paper. In particular, we consider a model with strategic workers, where the primary objective of a worker is to maximize his own utility. Based on this model, we analyze two basic mechanisms widely adopted in existing microtask crowdsourcing applications and show that, to obtain high quality solutions from workers, their costs are constrained by some lower bounds. We then propose a cost-effective mechanism that employs quality-aware worker training as a tool to stimulate workers to provide high quality solutions. We prove theoretically that the proposed mechanism, when properly designed, can obtain high quality solutions with an arbitrarily low cost. Beyond its theoretical guarantees, we further demonstrate the effectiveness of our proposed mechanisms through a set of behavioral experiments.
\end{abstract}


\newpage
\section{Introduction}
Crowdsourcing, which provides an innovative and effective way to access online labor market, has become increasingly important and prevalent in recent years. Until now, it has been successfully applied to a variety of applications ranging from challenging and creative projects such as R\&D challenges in InnoCentive [\ref{InnoCentive}] and software development tasks in TopCoder [\ref{TopCoder}], all the way to microtasks such as image tagging, keyword search and relevance feedback in Amazon Mechanical Turk (Mturk) [\ref{Mturk}] or Microworkers [\ref{Microworkers}]. Depending on the types of tasks, crowdsourcing takes different forms, which can be broadly divided into two categories: crowdsourcing contest and microtask crowdsourcing. Crowdsourcing contests are typically used for challenging and innovative tasks, where multiple workers simultaneously produce solutions to the same task for a requester who seeks and rewards only the highest-quality solution. On the other hand, microtask crowdsourcing targets on small tasks that are repetitive and tedious but easy for an individual to accomplish. Different from crowdsourcing contests, there exists no competition among workers in microtask crowdsourcing. In particular, workers will be paid a prescribed reward per task they complete, which is typically a small amount of money ranging from a few cents to a few dollars.

We focus on microtask crowdsourcing in this paper. With the access to large and relatively cheap online labor pool, microtask crowdsourcing has the advantage of solving large volumes of small tasks at a much lower price compared with traditional in-house solutions. However, due to the lack of proper incentives, microtask crowdsourcing suffers from quality issues. Since workers are paid a fixed amount of money per task they complete, it is profitable for them to provide random or bad quality solutions in order to increase the number of submissions within a certain amount of time or effort. It has been reported that most workers on Mturk, an leading marketplace for microtask crowdsourcing, do not contribute high quality work [\ref{wais}]. To make matters worse, there exists an inherent conflict between incentivizing high quality solutions from workers and maintaining the low cost advantage of microtask crowdsourcing for requesters. On the one hand, requesters typically have a very low budget for each task in microtask crowdsourcing. On the other hand, the implementation of incentive mechanisms is costly as the operation of verifying the quality of submitted solutions is expensive [\ref{Ipeirotis2010}]. Such a conflict makes it challenging to design incentives for microtask crowdsourcing, which motivates us to ask the following question: what incentive mechanisms should requesters employ to collect high quality solutions in a cost-effective way?

In this paper, we address this question from a game-theoretic perspective. In particular, we investigate a model with strategic workers, where the primary objective of a worker is to maximize his own utility, defined as the reward he will receive minus the cost of producing solutions of a certain quality. Based on this model, we first study two basic mechanisms widely adopted in existing microtask crowdsourcing applications. In particular, the first mechanism assigns the same task to multiple workers, identifies the correct solution for each task using a majority voting rule and rewards workers whose solution agrees with the correct one. The second mechanism assigns each task only to one worker, evaluates the quality of submitted solutions directly and rewards workers accordingly. We show that in order to obtain high quality solutions using these two mechanisms, the unit cost incurred by requesters per task is subject to a lower bound constraint, which is beyond the control of requesters and can be high enough to negate the low cost advantage of microtask crowdsourcing.

To tackle this challenge, we then propose a cost-effective mechanism that employs quality-aware worker training as a tool to stimulate workers to provide high quality solutions. In current microtask crowdsourcing applications, training tasks are usually assigned to workers at the very beginning and therefore is irrelevant to the quality of their submitted solutions. We show that assigning training tasks to workers when they perform poorly rather than uniformly at the beginning can effectively stimulate workers to produce high quality solutions. In particular, we prove theoretically that the proposed mechanism, when properly designed, can obtain high quality solutions with an arbitrarily low cost. Beyond its theoretical guarantees, we further conduct a serial of behavioral experiments to test our proposed mechanism. Our experimental results demonstrated the effectiveness of our proposed mechanism, and more generally the idea of quality-aware worker training, in stimulating high quality solutions at low costs.

The rest of the paper is organized as follows. Section II presents the related work. We introduce our model in Section III and study two basic mechanisms in Section IV. Then, in Section V, we describe the design of a cost-effective mechanism based on quality-aware worker training and analyze its performance. We show simulation results in Section VI and our experimental verifications in Section VII. Finally, we draw conclusions in Section VIII.

\section{Related Work}
Most of existing work on quality control for microtask crowdsourcing focuses on filtering and processing low quality submitted solutions [\ref{Ipeirotis2010}] - [\ref{Raykar2012}]. As oppose to such approaches, we study how to incentivize workers to produce high quality solutions in the first place. There has recently been work addressing incentives of crowdsourcing contests from game-theoretic perspectives by modeling these contests as all-pay auctions [\ref{DiPalantino09}] - [\ref{Cavallo12}]. Nevertheless, these models can not apply to our scenario as there exists no competition among workers in the context of microtask crowdsourcing.

There is a small literature that addresses incentives for microtask crowdsourcing. In [\ref{Shaw2010}], Shaw et al. conducted an experiment to compare the effectiveness of a collection of social and finical incentive mechanisms. In [\ref{Singer2011}], Singer and Mittal proposed a pricing scheme for microtask crowdsourcing where tasks are dynamically priced and allocated to workers based on their bids. A reputation-based incentive mechanism was proposed and analyzed for microtask crowdsourcing in [\ref{Zhang2012}]. Our work differs from these studies in that they do not consider the validation cost incurred by requesters in their models. For microtask crowdsourcing, the operation of verifying the quality of submitted solutions is so expensive that will negate its low cost advantage, which places a unique and practical challenge in the design of incentives. To the best of our knowledge, this is the first work that studies cost-effective incentive mechanisms for microtask crowdsourcing.

\section{The Model}
There are two main components in our model: requesters, who publish tasks; and workers, who produce solutions to the posted tasks. The submitted solutions can have varying quality, which is described by a one-dimensional value. Requesters maintain certain criteria on whether or not a submitted solution should be accepted. Only acceptable solutions are useful to requesters. Workers produce solutions to the posted tasks in return for reward provided by requesters. We assume workers are strategic, i.e., they choose the quality of their solutions selfishly to maximize their own utilities.

In our model, a mechanism describes how requesters evaluate the submitted solutions and reward workers accordingly. Mechanisms are designed by requesters with the aim of obtaining high quality solutions from workers, which should be published at the same time as tasks are posted. Mechanisms can be costly to requesters, which negates the advantages of crowdsourcing. In this work, we focus on mechanisms that not only can incentivize high quality solutions from workers, but also are cost-effective. We now formally describe the model.

\textbf{Worker Model.} We model the action of workers as the quality $q$ of their solutions. The value $q$ represents the probability of this solution is acceptable to requesters, which implies that $q \in [0,1]$. We assume that the solution space is infinite and the probability of two workers submitting the same unacceptable solution is 0. The cost incurred by a worker depends on the quality of solution he chooses to produce: a worker can produce a solution of quality $q$ at a cost $c(q)$. We make the following assumptions on the cost function $c(\cdot)$:
\begin{enumerate}
\item $c(q)$ is convex in $q$.
\item $c(q)$ is differentiable in $q$.
\item $c^{\prime}(q) > 0$, i.e., solutions with higher quality are more costly to produce.
\item $c(0) > 0$, i.e., even producing $0$ quality solutions will incur some cost.
\end{enumerate}

The benefit of a worker corresponds to the received reward, which depends on the quality of his solution, the mechanism being used and possibly the quality of other workers' solutions. We focus on symmetric scenarios, which means the benefit of a worker is evaluated under the assumption that all the other workers choose the same action (which may be different from the action of the worker under consideration). Denote by $V_{\mathcal{M}}(q,\tilde{q})$ the benefit of a worker who submits a solution of quality $q$ while other workers produce solutions with quality $\tilde{q}$ and mechanism $\mathcal{M}$ is employed by the requester. A quasi-linear utility is adopted, where the utility of a worker is the difference between his benefit and his cost:
\begin{equation}
u_{\mathcal{M}}(q,\tilde{q}) = V_{\mathcal{M}}(q,\tilde{q}) - c(q). \label{utility1}
\end{equation}

\textbf{Mechanism Choice.} Requesters employ mechanisms to incentivize high quality solutions from self-interested workers. Therefore, the action chosen by workers in response to a mechanism can be used to indicate the effectiveness of this mechanism. In particular, we will be interested in a desirable outcome where workers choose $q = 1$ as their optimal actions, i.e., self-interested workers are willing to contribute with the highest quality solutions. We would like to emphasize that such an outcome is practical in that microtasks are typically simple tasks that are easy for workers to accomplish satisfactorily. On the other hand, in a mechanism $\mathcal{M}$, there is a unit cost $C_{\mathcal{M}}$ per task incurred by the requester, which comes from the reward paid to workers and the cost for evaluating submitted solutions. We refer to such a unit cost $C_{\mathcal{M}}$ as the mechanism cost of $\mathcal{M}$. Since one of the main advantages of microtask crowdsourcing is its low cost, mechanisms should be designed to achieve the desirable outcome with low mechanism cost. Therefore, to study a certain mechanism, we wish to address the following questions: (a) under what conditions can we achieve the desirable outcome? and (b) what are the minimum mechanism cost and the corresponding parameter settings?

\textbf{Validation Approaches.} As an essential step towards incentivizing high quality solutions, a mechanism should be able to evaluate the quality of submitted solutions. We describe below three approaches considered in this paper, which are also commonly adopted in existing microtask crowdsourcing applications.

The first approach is majority voting, where requesters assign the same task to multiple workers and accept the solution that submitted by the majority of workers as the correct one. Clearly, the validation cost of majority voting depends on the number of workers per task. It has been reported that, if assigning the same task to more than 10 workers, the cost of microtask crowdsourcing solutions is comparable to that of in-house solutions [\ref{Ipeirotis2010}] and when the number of tasks is large, it is financially impractical to assign the same task to too many workers, e.g., more than 3 [\ref{wais}]. Therefore, when majority voting is adopted in incentive mechanisms, a key question need to be addressed: what is the minimum required number of workers per task to achieve the desirable outcome?

Second, requesters can use tasks with known solutions, which we refer to as gold standard tasks, to evaluate the submitted answers. Validation with gold standard tasks is expensive since correct answers are costly to obtain. More importantly, as the main objective of requesters in microtask crowdsourcing is to collect solutions for tasks, gold standard tasks can only be used occasionally for the purpose of assessing workers, e.g., as training tasks.

Note that both majority voting and gold standard tasks assume implicity that the task has a unique correct solution, which may not hold for creative tasks, e.g., writing a short description of a city. In this case, a quality control group [\ref{Hirth2012}] can be used to evaluate the submitted solution. In particularly, the quality group can be either a group of on-site experts who verify the quality of submitted solution manually or another group of workers who work on quality control tasks designed by the requesters. In the first case, the time and cost spent on evaluating the submitted solutions is typically comparable to that of performing the task itself. In the second case, requesters not only have to investigate time and effort in designing quality control tasks but also need to pay workers for working these tasks. Therefore, validation using quality control group is also an expensive operation.

\section{Basic Incentive Mechanisms}
We study in this section two basic mechanisms that are widely employed in existing microtask crowdsourcing applications. Particularly, for each mechanism, we characterize conditions under which workers will choose $q = 1$ as their best responses and study the minimum mechanism cost for achieving it.

\subsection{A Reward Consensus Mechanism}
\label{MDMechanism}

We first consider a mechanism that employs majority voting as its validation approach and, when a consensus is reached, rewards workers who submitted the consensus solution. We refer to such a mechanism as the reward consensus mechanism and denote it by $\mathcal{M}_c$. In $\mathcal{M}_c$, a task is assigned to $K + 1$ different workers. We assume that $K$ is an even number and is greater than $0$. If the same solution is submitted by no less than $K/2 + 1$ workers, then it is chosen as the correct solution. Workers are paid the prescribed reward $r$ if they submit the correct solution. On the other hand, workers will receive no payments if their submitted solutions are different from the correct one or if no correct solution can be identified, i.e., no consensus is reached.

In $\mathcal{M}_c$, the benefit of each worker depends not only on his own action but also on other workers' actions. Therefore, a worker will condition his decision making on others' actions, which results in couplings in workers' actions. To capture such interactions among workers, we adopt the solution concept of symmetric Nash equilibrium, which can be formally stated as:

\begin{definition}[Symmetric Nash Equilibrium of $\mathcal{M}_c$] \label{definition1}
The $q^*$ is a symmetric Nash equilibrium in $\mathcal{M}_c$ if $q^*$ is the best response of a worker when other workers are choosing $q^*$.
\end{definition}

We show below the necessary and sufficient conditions of $q^* = 1$ being a symmetric Nash equilibrium in $\mathcal{M}_c$.
\begin{proposition}
In $\mathcal{M}_c$, $q^* = 1$ is a symmetric Nash equilibrium if and only if $r \ge c^{\prime}(1)$.
\end{proposition}
\begin{IEEEproof}
Under the assumption that the probability of any two workers submitting the same unacceptable solution is zero (which is reasonable as there are infinitely possible solutions), we can calculate the utility of an worker who produces solutions of quality $q$ while other workers choose action $\tilde{q}$ as
\begin{equation}
u_{\mathcal{M}_c}(q,\tilde{q}) = r q \sum\limits_{n = K/2}^K \frac{K!}{n!(K - n)!} \tilde{q}^n(1 - \tilde{q})^{K - n} - c(q). \nonumber
\end{equation}

According to Definition \ref{definition1}, $q^*$ is a symmetric Nash Equilibrium of $\mathcal{M}_c$ if and only if
\begin{equation}
q^* \in \arg\max_{q \in [0,1]} u_{\mathcal{M}_c}(q,q^*).
\end{equation}
Since $u_{\mathcal{M}_c}(q,1) = r q - c(q)$ is a concave function of $q$ and $q \in [0, 1]$, the necessary and sufficient condition of $q^* = 1$ being a symmetric Nash equilibrium can be derived as
\begin{equation}
\frac{\partial u_{\mathcal{M}_c}(q,1)}{\partial q}|_{q = 1} = r - c^{\prime}(1) \ge 0.
\end{equation}
\end{IEEEproof}

From Proposition 1, we can see that $\mathcal{M}_c$ can enforce self-interested workers to produce the highest quality solutions as long as the prescribed reward $r$ is larger than a certain threshold. Surprisingly, this threshold depends purely on the worker's cost function and is irrelevant to the number of workers. The mechanism cost of $\mathcal{M}_c$ can be calculated as
\begin{equation}
C_{\mathcal{M}_c} = (K + 1)r \ge (K + 1)c^{\prime}(1).
\end{equation}
Therefore, to minimize the mechanism cost, it is optimal to choose the minimum value of $K$, i.e., $K = 2$, and let $r = c^{\prime}(1)$. In this way, requesters can achieve $q^* = 1$ with the minimum mechanism cost $C^*_{\mathcal{M}_c} = 3c^{\prime}(1)$. Having more workers working on the same task will only increase the mechanism cost while not helping to improve the quality of submitted solutions.

\subsection{A Reward Accuracy Mechanism}

Next, we consider a mechanism that rewards a worker purely based on his own submitted solutions. Such a mechanism is referred to as the reward accuracy mechanism and is denoted by $\mathcal{M}_a$. In particular, depending on the characteristics of tasks, $\mathcal{M}_a$ will use either gold standard tasks or the quality control group to verify whether a submitted solution is acceptable or not. In our discussions, however, we make no distinctions between the two methods. We assume that the validation cost per task is $d$ and there is a certain probability $\epsilon \ll 1$ that a mistake will be made in deciding whether a solution is acceptable or not.

As we have discussed, these validation operations are expensive and should be used rarely. Therefore, $\mathcal{M}_a$ only evaluates randomly a fraction of submitted solutions to reduce the mechanism cost. Formally, in $\mathcal{M}_a$, requesters verify a submitted solution with probability $\alpha_a$. If a submitted solution is acceptable or not evaluated, the worker will receive the prescribed reward $r$. On the other hand, if the solution being evaluated is unacceptable, the worker will not be paid.

In $\mathcal{M}_a$, the utility of a worker is irrelevant to actions of other workers. Therefore, we write the utility of a worker who produces solutions of quality $q$ as
\begin{equation}
u_{\mathcal{M}_a}(q) = r\left[(1 - \alpha_a) + \alpha_a (1 - \epsilon)q + \alpha_a \epsilon (1 - q) \right] - c(q). \nonumber
\end{equation}

Let $q^*$ represent the optimal action of a worker by which his utility function is maximized. Since $u_{\mathcal{M}_a}(q)$ is a concave function of $q$ and $q \in [0, 1]$, we can derive the necessary and sufficient conditions of $q^* = 1$ as
\begin{equation}
\alpha_a \ge \frac{c^{\prime}(1)}{(1 - 2\epsilon)r}.
\end{equation}

We can see that there is a lower bound on possible values of $\alpha_a$, which depends on the cost function of workers and the prescribed reward $r$. Since $\alpha_a \in [0, 1]$, for the above condition to hold, we must have $r \ge \frac{c^{\prime}(1)}{(1 - 2\epsilon)}$.  Moreover, we can calculate the mechanism cost in the case of $q^* = 1$ as
\begin{equation}
C_{\mathcal{M}_a} = (1 - \alpha_a\epsilon)r + \alpha_a d.  \nonumber
\end{equation}

Requesters optimize the mechanism cost by choosing the sampling probability $\alpha_a$ and the reward $r$. Therefore, we can calculate the minimum mechanism cost as
\begin{equation}
C^*_{\mathcal{M}_a} = \min\limits_{\frac{c^{\prime}(1)}{(1 - 2\epsilon)r} \le \alpha_a \le 1, \ r\ge \frac{c^{\prime}(1)}{(1 - 2\epsilon)}} (1 - \alpha_a\epsilon)r + \alpha_a d.
\end{equation}

By solving the above convex optimization problem using the Karush-Kuhn-Tucker conditions [\ref{convex}], we get
\begin{equation}
C^*_{\mathcal{M}_a} = \left\{ {\begin{array}{*{20}{l}}
&2\sqrt{\frac{c^{\prime}(1)d}{1 - 2\epsilon}} - \epsilon \frac{c^{\prime}(1)}{1 - 2\epsilon}, &\text{          if      } d \ge \frac{c^{\prime}(1)}{1 - 2\epsilon},  \\
&\frac{c^{\prime}(1)(1 - \epsilon)}{1 - 2\epsilon} + d, &\text{           otherwise}.
\end{array}} \right.
\end{equation}
Moreover, the optimal parameters for achieving the minimum mechanism cost are
\begin{equation}
\left\{ {\begin{array}{*{20}{l}}
&\alpha^*_a = \sqrt{\frac{c^{\prime}(1)}{(1 - 2\epsilon})d},\  r^* = \sqrt{\frac{c^{\prime}(1)d}{1 - 2\epsilon}}, &\text{                if      } d \ge \frac{c^{\prime}(1)}{1 - 2\epsilon},\\
&\alpha^*_a = 1, \ r^* = \frac{c^{\prime}(1)}{1 - 2\epsilon}, &\text{                         otherwise}.
\end{array}} \right.
\end{equation}

Similarly as the reward consensus mechanism, the mechanism cost of the reward accuracy mechanism must be greater than a certain value in order for requesters to collect solutions with the highest quality from workers.

\section{Reducing Mechanism Cost By Quality-Aware Worker Training}

From our discussions above, we can see that for the two basic mechanisms to achieve the desirable outcome, their mechanism costs are constrained by some lower bounds, i.e., the minimum mechanism costs. These minimum mechanism costs are determined by worker's cost function and possibly the validation cost, all of which are beyond the control of requesters. If these minimum mechanism costs are large, requesters will have to either lower their standards and suffer from low quality solutions or switch to other alternative approaches.

To overcome this issue, we introduce a new mechanism $\mathcal{M}_t$, which employs quality-aware worker training as a tool to stimulate self-interested workers to submit high quality solutions. Our proposed mechanism is built on top of the basic mechanisms to further reduce the required mechanism cost. In particular, there are two states in $\mathcal{M}_t$: the working state, where workers work on standard tasks in return for reward; and the training state, where workers do a set of training tasks to gain qualifications for the working state.

In the working state, we consider a general model which incorporates both the reward consensus mechanism and the reward accuracy mechanism. We assume that with probability $1 - \beta_w$, a task will go through the reward consensus mechanism and with probability $\beta_w$, the reward accuracy mechanism will be used with the sampling probability $\alpha_w$. According to our results in Section \ref{MDMechanism}, it is optimal to assign $3$ workers per task when the reward consensus mechanism is being used. In the working state, a submitted solution will be accepted by $\mathcal{M}_t$ if it is accepted by either the reward consensus mechanism or the reward accuracy mechanism. A submitted solution will be rejected otherwise. When a solution is accepted, the worker will receive the prescribed reward $r$ and can continue working on more tasks in the working state. On the other hand, if a worker's solution is rejected, he will not be paid for this task and will be put into the training state to earn his qualifications for future tasks. Let $P_w(\tilde{q}_w, q_w)$ represent the probability of a solution with quality $q_w$ being accepted in the working state when other submitted solutions are of quality $\tilde{q}_w$. We have
\begin{align}
P_w(\tilde{q}_w, q_w) = & (1 - \beta_w) q_w \left[\tilde{q}^2_w + 2 \tilde{q}_w (1 - \tilde{q}_w)\right] + \beta_w(1 - \alpha_w) + \beta_w \alpha_w[(1 - 2\epsilon)q_w + \epsilon]. \label{tranProb1}
\end{align}
The immediate utility of a worker at the working state can be calculated as
\begin{equation}
u^{w}_{\mathcal{M}_t}(\tilde{q}_w, q_w) = r P_w(\tilde{q}_w, q_w) - c(q_w). \label{immeUtil1}
\end{equation}

In the training state, each worker will receive a set of $N$ training tasks. To evaluate the submitted solutions, an approach similar to the reward accuracy mechanism is adopted. In particular, a worker is chosen to be evaluated at random with probability $\alpha_t$. A chosen worker will pass the evaluation and gain the permission to working state if $M$ out $N$ solutions are correct. We assume $M = N$ in our analysis while our results can be easily extended to more general cases. An unselected worker will be granted the permission to working state next time. Only workers who fail the evaluation will stay in the training state and receive another set of $N$ training tasks. We denote by $P_t(q_t)$ the probability of a worker who produces solutions of quality $q_t$ being allowed to enter the working state next time, which can be calculated as
\begin{equation}
P_t(q_t) = (1 - \alpha_t) + \alpha_t [(1 - 2 \epsilon)q_t + \epsilon]^N. \label{tranProb2}
\end{equation}
The immediate utility of a worker at the training state is
\begin{equation}
u^{t}_{\mathcal{M}_t}(q_t) = - N c(q_t). \label{immeUtil2}
\end{equation}

\begin{figure}
\centering
\includegraphics[width=4.0in]{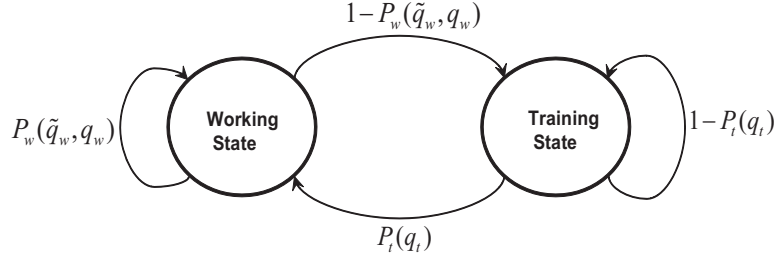}
\caption{The state transition diagram of our proposed mechanism $\mathcal{M}_t$.}
\end{figure}

To summarize, we plot the state transitions of $\mathcal{M}_t$ in Fig. 1. We further assume that at the end of each time slot, a worker will leave the system with probability $1 - \delta$, where $\delta \in (0,1)$. Moreover, a new worker will enter the system immediately after an existing one left. New workers will be placed randomly into the working state or the training state according to an initial state distribution specified by the requester.

From (\ref{immeUtil1}) and (\ref{immeUtil2}), we can see that workers' immediate utility in $\mathcal{M}_t$ depends not only on their actions but also on which state they are in. Moreover, as the state transition probabilities depend on workers' actions according to (\ref{tranProb1}) and (\ref{tranProb2}), taking a certain action will affect not only the immediate utility but also the future utility. For example, a worker may increase his immediate utility by submitting poor solutions at the working state but suffer from the loss of being placed into the training state next time. Given the dependence of future utility on current actions, as rational decision makers, workers will choose their actions to optimize their long-term utility. Formally, we denote by $U^{w}_{\mathcal{M}_t}(\tilde{q}_w, q_w, q_t)$ the long-term expected utility of a worker who is currently at the working state and chooses action $q_w$ for the working state and action $q_t$ for the training state while others choosing action $\tilde{q}_w$ at the working state. Similarly, we write $U^{t}_{\mathcal{M}_t}(\tilde{q}_w, q_w, q_t)$ for the long-term expected utility at the training state. We have
\begin{eqnarray}
\!\!\!\!\!\!\!\!U^{w}_{\mathcal{M}_t}(\tilde{q}_w, q_w, q_t) \!\!\!\!&=&\!\!\!\! u^{w}_{\mathcal{M}_t}(\tilde{q}_w, q_w) \!+\! \delta\!\left[P_w(\tilde{q}_w, q_w) U^{w}_{\mathcal{M}_t}(\tilde{q}_w, q_w, q_t) \!+\! (1 \!-\! P_w(\tilde{q}_w, q_w)) U^{t}_{\mathcal{M}_t}(\tilde{q}_w, q_w, q_t)\right], \ \ \ \ \ \ \label{longtermU11}\\
\!\!\!\!\!\!\!\!U^{t}_{\mathcal{M}_t}(\tilde{q}_w, q_w, q_t) \!\!\!\!&=&\!\!\!\! u^{t}_{\mathcal{M}_t}(q_t) + \delta\left[P_t(q_t) U^{w}_{\mathcal{M}_t}(\tilde{q}_w, q_w, q_t) + (1 - P_t(q_t)) U^{t}_{\mathcal{M}_t}(\tilde{q}_w, q_w, q_t)\right]. \label{longtermU12}
\end{eqnarray}

Based on the definition of worker's long-term expected utility, we adopt the symmetric Nash equilibrium as the solution concept in mechanism $\mathcal{M}_t$, which is formally defined as

\begin{definition}[Symmetric Nash Equilibrium of $\mathcal{M}_t$] \label{definition2}
The action pair $(\hat{q}_w, \hat{q}_t)$ is a symmetric Nash equilibrium of $\mathcal{M}_t$, if $\forall q_w \in [0,1]$ and $\forall q_t \in [0,1]$, the following two conditions hold
\begin{eqnarray}
U^{w}_{\mathcal{M}_t}(\hat{q}_w, \hat{q}_w, \hat{q}_t) \ge U^{w}_{\mathcal{M}_t}(\hat{q}_w, q_w, q_t), \label{MDP11}  \\
U^{t}_{\mathcal{M}_t}(\hat{q}_w, \hat{q}_w, \hat{q}_t) \ge  U^{t}_{\mathcal{M}_t}(\hat{q}_w, q_w, q_t). \label{MDP12}
\end{eqnarray}
\end{definition}

The above definition suggests a way to verify whether an action pair $(\hat{q}_w, \hat{q}_t)$ of interest is a symmetric Nash equilibrium or not, which can be summarized as the following three steps.
\begin{enumerate}
\item Assume all workers are adopting $(\hat{q}_w, \hat{q}_t)$ and one worker of interest may deviate from it.
\item Find the optimal action $(q^*_w, q^*_t)$ for this worker.
\item The action pair $(\hat{q}_w, \hat{q}_t)$ is a symmetric Nash equilibrium if and only if it is consistent with the optimal action pair $(q^*_w, q^*_t)$, i.e., $\hat{q}_w = q^*_w$ and $\hat{q}_t = q^*_t$.
\end{enumerate}

The key challenge here is to find the optimal action pair for a worker given the other workers' action, which can be modeled as a Markov Decision Process (MDP). In this MDP formulation, the state set includes the working state and the training state, the action in each state is the quality of solutions to produce, rewards are the immediate utility specified in (\ref{immeUtil1}) and (\ref{immeUtil2}), and transition probabilities are given in (\ref{tranProb1}) and (\ref{tranProb2}).

Note that in our discussions so far we assume stationary actions, i.e., workers' actions are time-invariant functions of the state. Such an assumption can be justified by properties of MDP as shown in Proposition 2.
\begin{proposition}
Any worker cannot improve his long-term expected utility by choosing time-variant actions, if all the other workers' action at the working state is stationary, i.e., $\forall q_w \in [0,1]$,
\begin{eqnarray}
U^{w}_{\mathcal{M}_t}(q_w, q^*_w(\tau), q^*_t(\tau)) &=& U^{w}_{\mathcal{M}_t}(q_w, q^*_w, q^*_t), \nonumber\\
U^{t}_{\mathcal{M}_t}(q_w, q^*_w(\tau), q^*_t(\tau)) &=& U^{t}_{\mathcal{M}_t}(q_w, q^*_w, q^*_t), \nonumber
\end{eqnarray}
where $(q^*_w(\tau), q^*_t(\tau))$ is the optimal time-variant action pair and $(q^*_w, q^*_t)$ is the optimal stationary action pair, given other workers' action $q_w$.
\end{proposition}
\begin{IEEEproof}
The problem of finding the optimal action pair for a worker given the other workers' action can be formulated as a MDP. In this MDP formulation, rewards and transition probabilities are stationary if other workers' action at the working state is stationary. In addition, the state space is stationary and finite and the action space is stationary and compact. Moreover, the rewards and transition probabilities are continuous in actions. Therefore, according to Theorem 6.2.10 in [\ref{mdp}], there exits a deterministic stationary action rule by which the optimal utility of this MDP can be achieved. In other words, choosing any random, time-variant and history dependent action rules will not lead to a higher utility.
\end{IEEEproof}

Among all possible symmetric Nash equilibria, we are interested in ones where $\hat{q}_w = 1$, i.e., workers will produce solutions with the highest quality at the working state. Note that we do not guarantee solution quality at the training state since in $\mathcal{M}_t$, the working state serves the production purpose whereas the training state is designed as an auxiliary state to enhance workers' performance at the working state. Solutions collected from the training state will only be used for assessing workers and should be discarded afterwards. We would like to characterize conditions under which such symmetric Nash equilibria exist. Toward this end, we will follow the three steps outlined above with an emphasize on solving the MDP to find the optimal action pair. Our results are summarized in the following proposition, where we present a necessary and sufficient condition on the existence of symmetric Nash equilibria with $\hat{q}_w = 1$.

\begin{proposition}
There exists $\hat{q}_t \in [0,1]$ such that $(1, \hat{q}_t)$ is a symmetric Nash equilibrium of $\mathcal{M}_t$ if and only if
\begin{equation}
U^{w}_{\mathcal{M}_t}(1, 1, \hat{q}_t) - U^{t}_{\mathcal{M}_t}(1, 1, \hat{q}_t) \ge \frac{c^{\prime}(1)}{\delta \left[(1 - \beta_w) + \beta_w\alpha_w (1 - 2 \epsilon)\right]} - \frac{r}{\delta}. \label{NScondition1}
\end{equation}
\end{proposition}
\begin{IEEEproof}
To show the existence of a symmetric Nash equilibrium with $\hat{q}_w = 1$, we first assume that all workers are choosing the action pair $(1, \hat{q}_t)$ except one worker under consideration. Since interactions among workers only occur at the working state, the value of $\hat{q}_t$ will not affect the decision of this particular worker.

Next, we characterize the optimal action pair $(q^*_w,q^*_t)$ for this particular worker. The problem of finding the optimal action pair of a certain worker can be modeled as a MDP where the necessary and sufficient conditions of an action pair being optimal are given in (\ref{MDP11}) and (\ref{MDP12}). Nevertheless, it is not easy to derive the optimal action pair directly from these conditions. Therefore, we need to find another set of equivalent conditions. Since in our MDP formulation, $0<\delta<1$, the state space is finite and the immediate reward is bounded, Theorem 6.2.7 in [\ref{mdp}] shows that an action pair $(q^*_w,q^*_t)$ is optimal if and only if it satisfies the following optimality equations
\begin{eqnarray}
\!\!\!\!q^*_w \!\!\!\!&\in&\!\!\!\! \arg\max_{0\le q_w \le 1} \left\{  u^{w}_{\mathcal{M}_t}(1, q_w) \!+\! \delta\!\left[P_w(1, q_w) U^{w}_{\mathcal{M}_t}(1, q^*_w, q^*_t) \!+\! (1 \!-\! P_w(1, q_w)) U^{t}_{\mathcal{M}_t}(1, q^*_w, q^*_t)\right] \right\}, \ \  \label{optEqn1}\\
\!\!\!\!q^*_t \!\!\!\!&\in&\!\!\!\! \arg\max_{0\le q_t \le 1} \left\{  u^{t}_{\mathcal{M}_t}(q_t) + \delta\left[P_t(q_t) U^{w}_{\mathcal{M}_t}(1, q^*_w, q^*_t) + (1 - P_t(q_t)) U^{t}_{\mathcal{M}_t}(1, q^*_w, q^*_t) \right] \right\}, \label{optEqn2}
\end{eqnarray}
and that there exits at least one optimal action pair.

Since the above optimality equations hold for any value of $\hat{q}_t$, we set $\hat{q}_t = q^*_t$. Then, to prove that there exists an symmetric Nash equilibrium $(\hat{q}_w, \hat{q}_t)$ with $\hat{q}_w = 1$, it suffices to show that $q^*_w = 1$. Substituting (\ref{immeUtil1}) into (\ref{optEqn1}) and after some manipulations, we have
\begin{equation}
q^*_w \in \arg\max_{0\le q_w \le 1} \left\{  \left[r + \delta U^{w}_{\mathcal{M}_t}(1, q^*_w, q^*_t) - \delta U^{t}_{\mathcal{M}_t}(1, q^*_w, q^*_t)\right] P_w(1, q_w) - c(q_w) \right\}. \label{OptimalAction1}
\end{equation}
From (\ref{tranProb1}), we know
\begin{equation}
P_w(1, q_w) = \left[ (1 - \beta_w) + \beta_w\alpha_w (1 - 2 \epsilon)\right] q_w + \beta_w(1 - \alpha_w) + \beta_w\alpha_w \epsilon. \label{Pw1}
\end{equation}
Substituting (\ref{Pw1}) into (\ref{OptimalAction1}), we have
\begin{equation}
q^*_w \in \arg\max_{0\le q_w \le 1} \left\{  \left[(1 - \beta_w) + \beta_w\alpha_w (1 - 2 \epsilon)\right] \left[r + \delta U^{w}_{\mathcal{M}_t}(1, q^*_w, q^*_t) - \delta U^{t}_{\mathcal{M}_t}(1, q^*_w, q^*_t)\right] q_w - c(q_w) \right\}.  \nonumber
\end{equation}

Recall that $c(q_w)$ is a convex function of $q_w$. We can thus derive the necessary and sufficient condition for $q^*_w = 1$ as
\begin{equation}
\left[ (1 - \beta_w) + \beta_w\alpha_w (1 - 2 \epsilon)\right] \left[r + \delta U^{w}_{\mathcal{M}_t}(1, 1, q^*_t) - \delta U^{t}_{\mathcal{M}_t}(1, 1, q^*_t)\right] \ge c^{\prime}(1),
\end{equation}
which is also the necessary and sufficient condition for the existence of the symmetric Nash equilibrium $(\hat{q}_w, \hat{q}_t)$ with $\hat{q}_w = 1$. Replacing $q^*_t$ with $\hat{q}_t$, we obtain the condition in (\ref{NScondition1}) and complete the proof.
\end{IEEEproof}

In the above proposition, we show that it is an equilibrium for self-interested workers to produce solutions with quality $1$ at the working state as long as the condition in (\ref{NScondition1}) holds. Nevertheless, this condition is hard to evaluate since neither the equilibrium action at the training state, $\hat{q}_t$, nor the optimal long-term utility $U^{w}_{\mathcal{M}_t}(1, 1, \hat{q}_t)$ and $U^{t}_{\mathcal{M}_t}(1, 1, \hat{q}_t)$ are known to requesters. On the other hand, we hope to find conditions that can provide guide requesters in choosing proper parameters for mechanism $\mathcal{M}_t$. Therefore, based on results of Proposition 3, we present in the following a sufficient condition on the existence of desirable equilibria, which is also easy to evaluate.

\begin{theorem}
In $\mathcal{M}_t$, if the number of training tasks $N$ is large enough, i.e.,
\begin{equation}
\label{suffCond}
N \ge \frac{1}{c(0)} \left[\frac{(1 + \delta\beta_w\alpha_w \epsilon)c^{\prime}(1)}{\delta (1 - \beta_w) + \delta \beta_w\alpha_w (1 - 2 \epsilon)} - \frac{\delta + 1}{\delta} r + c(1)\right],
\end{equation}
then there exits a symmetric Nash equilibrium $(\hat{q}_w, \hat{q}_t)$ such that $\hat{q}_w = 1$.
\end{theorem}
\begin{IEEEproof} We first obtain a lower bound on $U^{w}_{\mathcal{M}_t}(1, 1, \hat{q}_t) - U^{t}_{\mathcal{M}_t}(1, 1, \hat{q}_t)$ and then combine this lower bound with Proposition 3 to prove Theorem 1.

Let $\mathbf{U}(q_w, q_t) \triangleq \left[U^{w}_{\mathcal{M}_t}(1, q_w, q_t)\ \  U^{t}_{\mathcal{M}_t}(1, q_w, q_t)\right]^T$. Then, from (\ref{longtermU11}) and (\ref{longtermU12}), we have
\begin{equation}
\left(\mathbf{I} - \delta \mathbf{Q}(q_w, q_t)\right)\mathbf{U}(q_w, q_t) = \mathbf{b}(q_w, q_t),
\end{equation}
where $\mathbf{I}$ is a 2 by 2 identity matrix, $\mathbf{b}(q_w, q_t) \triangleq [u^{w}_{\mathcal{M}_t}(1, q_w)\ \  u^{t}_{\mathcal{M}_t}(q_t)]^T$ and
\begin{equation}
\mathbf{Q}(q_w, q_t) \triangleq \left[ {\begin{array}{*{20}{c}}
{{P_w}(1,{q_w})}&{1 - {P_w}(1,{q_w})}\\
{{P_t}({q_t})}&{1 - {P_t}({q_t})}
\end{array}} \right].
\end{equation}

Since $0 < \delta < 1$, it can be proved according to the Corollary C.4 in [\ref{mdp}] that matrix $(\mathbf{I} - \delta \mathbf{Q}(q_w, q_t))$ is invertible. Therefore, we can obtain the long-term utility vector of action pair $(q_w, q_t)$ as
\begin{equation}
\mathbf{U}(q_w, q_t) = \left(\mathbf{I} - \delta \mathbf{Q}(q_w, q_t)\right)^{-1} \mathbf{b}(q_w, q_t). \label{longtermU2}
\end{equation}

Based on (\ref{longtermU2}), we have
\begin{eqnarray}
U^{w}_{\mathcal{M}_t}(1, q_w, q_t) - U^{t}_{\mathcal{M}_t}(1, q_w, q_t) &=& [1 \  -\!1]\mathbf{U}(q_w, q_t) \nonumber \\
&=& \frac{u^{w}_{\mathcal{M}_t}(1, q_w) - u^{t}_{\mathcal{M}_t}(q_t)}{1 + \delta \left[P_t(q_t) - P_w(1,q_w) \right]}.
\end{eqnarray}

The above results hold for $\forall q_w \in [0, 1]$ and $\forall q_t \in [0, 1]$. Therefore, for a desired action pair $(1, \hat{q}_t)$, we have
\begin{eqnarray}
U^{w}_{\mathcal{M}_t}(1, 1, \hat{q}_t) - U^{t}_{\mathcal{M}_t}(1, 1, \hat{q}_t) &=& \frac{u^{w}_{\mathcal{M}_t}(1, 1) - u^{t}_{\mathcal{M}_t}(\hat{q}_t)}{1 + \delta \left[P_t(\hat{q}_t) - P_w(1,1) \right]} \nonumber \\
&=& \frac{(1 - \beta_w\alpha_w \epsilon)r - c(1) + N c(\hat{q}_t)}{1 + \delta\left\{1 - \alpha_t + \alpha_t [(1 - 2 \epsilon)\hat{q}_t + \epsilon]^N - (1 - \beta_w\alpha_w \epsilon)\right\}} \nonumber \\
&\ge & \frac{(1 - \beta_w\alpha_w \epsilon)r - c(1) + N c(0)}{1 + \delta\beta_w\alpha_w \epsilon}. \label{lowerBound1}
\end{eqnarray}
Since $[(1 - 2 \epsilon)\hat{q}_t + \epsilon]^N \le 1$, the inequality in (\ref{lowerBound1}) is derived by replacing $[(1 - 2 \epsilon)\hat{q}_t + \epsilon]^N$ with 1 and by using the fact that $c(q)$ is monotonically increasing in $q$.

Therefore, the condition in (\ref{NScondition1}) is guaranteed to hold if
\begin{equation}
\label{suffCond0}
\frac{(1 - \beta_w\alpha_w \epsilon)r - c(1) + N c(0)}{1 + \delta\beta_w\alpha_w \epsilon} \ge \frac{c^{\prime}(1)}{\delta \left[ (1 - \beta_w) + \beta_w\alpha_w (1 - 2 \epsilon)\right]} - \frac{r}{\delta}, \nonumber
\end{equation}
which leads to the sufficient condition in (\ref{suffCond}).
\end{IEEEproof}

Theorem 1 shows that given any possible settings $(\alpha_w, \beta_w, r, \alpha_t)$ in $\mathcal{M}_t$, we can always enforce workers to produce solutions with quality $1$ at the working state by choosing a sufficiently large $N$. Such a property makes it possible for requesters to control their cost while obtaining high quality solutions. We discuss the mechanism cost of $\mathcal{M}_t$ in the following subsection.
\subsection{Mechanism Cost}
For requesters, the mechanism cost of $\mathcal{M}_t$ at the desirable equilibrium $(1, \hat{q}_t)$ can be written as
\begin{equation}
C_{\mathcal{M}_t} = (1 - \beta_w)\cdot 3r + \beta_w\cdot\left[(1-\alpha_w\epsilon)r + \alpha_w d \right] + \beta_w\cdot\alpha_w\epsilon\sum\limits_{k=0}^{\infty}\left[1 - P_t(\hat{q}_t)\right]^k \alpha_t N d,
\nonumber
\end{equation}
where the last term corresponds to the cost of validation in the training state. Since $\epsilon \ll 1$, it follows that $P_t(\hat{q}_t) \ge 1 - \alpha_t + \alpha_t \epsilon^N$. Therefore, we have
\begin{equation}
C_{\mathcal{M}_t} \le 3r(1 - \beta_w) + \beta_w\left[(1-\alpha_w\epsilon)r + \alpha_w d \right] + \frac{\alpha_t}{1 - \alpha_t( 1 - \epsilon^N)}\beta_w\alpha_w\epsilon N d . \nonumber
\end{equation}

We then design parameters of $\mathcal{M}_t$ according to the following procedure: (a) select working state parameters $\alpha_w$, $\beta_w$ and $r$, (b) choose $N$ such that (\ref{lowerBound1}) holds, (c) design $\alpha_t$ such that
\begin{equation}
\frac{\alpha_t}{1 - \alpha_t( 1 - \epsilon^N)}\beta_w\alpha_w\epsilon N d \le \gamma \{3r(1 - \beta_w) + \beta_w\left[(1-\alpha_w\epsilon)r + \alpha_w d \right]\}, \label{alpha_t}
\end{equation}
where $\gamma > 0$ is a parameter chosen by requesters to control the relative cost of training state to working state. The inequality in (\ref{alpha_t}) is equivalent to
\begin{equation}
\alpha_t \le \frac{\gamma \{3r(1 - \beta_w) + \beta_w\left[(1-\alpha_w\epsilon)r + \alpha_w d \right]\} }{\gamma (1 - \epsilon^N) \{3r(1 - \beta_w) + \beta_w\left[(1-\alpha_w\epsilon)r + \alpha_w d \right]\} + \beta_w\alpha_w\epsilon N d}. \label{alpha_t2}
\end{equation}

Following the above design procedure, we have
\begin{equation}
C_{\mathcal{M}_t} \le (1 + \gamma)\left[3r(1 - \beta_w) + \beta_w((1-\alpha_w\epsilon)r + \alpha_w d)\right]. \nonumber
\end{equation}
If $\alpha_w$ and $r$ are chosen to minimize the cost, we have
\begin{equation}
C^*_{\mathcal{M}_t} = \inf\limits_{0 < \alpha_w \le 1, r > 0} (1 + \gamma)\left[3r(1 - \beta_w) + \beta_w((1-\alpha_w\epsilon)r + \alpha_w d)\right] = 0, \nonumber
\end{equation}
which illustrates that our proposed mechanism $\mathcal{M}_t$ enables requesters to obtain high quality solutions with an arbitrarily low cost.

Moreover, from the above design procedure, the significance of our proposed mechanism can be interpreted from another perspective. That is, through the introduction of quality-aware worker training, our proposed mechanism can be built on top of any basic mechanisms to bring requesters an extra degree of freedom in their design. They can now freely choose working state parameters, e.g., $\alpha_w$, $\beta_w$ and $r$, without concerning the constraint of incentivizing high quality solutions, which will be automatically guaranteed through the design of training state parameters.

\subsection{Stationary State Distribution}
In above discussions, we focus on the quality of submitted solutions at the working state, while there is no guarantee of solution quality at the training state. This is sufficient for requesters to high quality solutions as the training state only serves as an axillary state and will not be used for production. On the other hand, the system efficiency of $\mathcal{M}_t$ depends on the probability of a worker being at the working state. If such a probability is small, $\mathcal{M}_t$ will have low efficiency as a large portion of workers are not contributing to actual tasks

Therefore, to fully study the performance of $\mathcal{M}_t$, we analyze the stationary state distribution of $\mathcal{M}_t$ in this subsection. We denote by $\pi^n_w$ the probability of a worker being at the working state at the $n$th time slot after entering the platform. The probability of being at the training state is thus $(1 - \pi^n_w)$. We denote by $\pi^{\infty}_w$ and $\pi^0_w$ the stationary state distribution and initial state distribution, respectively. Note that the initial state distribution $\pi^0_w$ is a design aspect that can be controlled by requesters, i.e., requesters can decide whether a new worker starts at the working state or at the training state. Our main result is a lower bound of $\pi^{\infty}_w$ as shown in the following proposition.

\begin{proposition}
In $\mathcal{M}_t$, if workers follow a desirable symmetric Nash equilibrium $(1, \hat{q}_t)$, then the stationary state distribution $\pi^{\infty}_w$ will be reached and
\begin{equation}
\pi^{\infty}_w \ge \frac{(1 - \delta)\pi^{0}_w + \delta (1 - \alpha_t)}{1 - \delta + \delta \beta_w \alpha_w \epsilon + \delta(1 - \alpha_t)}
\end{equation}
\end{proposition}
\begin{IEEEproof}
Assuming that all workers are adopting the action pair $(1, \hat{q}_t)$, then we can write the state distribution update rule as
\begin{eqnarray}
\pi^{n + 1}_w &=& \delta \pi^{n}_w P_w(1,1) + \delta (1 - \pi^{n}_w)P_t(\hat{q}_t) + (1 - \delta)\pi^{0}_w \nonumber \\
&=& \delta\left[P_w(1,1) - P_t(\hat{q}_t)\right]\pi^{n}_w + (1 - \delta)\pi^{0}_w + \delta P_t(\hat{q}_t). \label{stateDistUpdate}
\end{eqnarray}
If the stationary state distribution $\pi^{\infty}_w$ exists, it must satisfy
\begin{equation}
\pi^{\infty}_w = \delta\left[P_w(1,1) - P_t(\hat{q}_t)\right]\pi^{\infty}_w + (1 - \delta)\pi^{0}_w + \delta P_t(\hat{q}_t). \label{stationaryStateCond1}
\end{equation}
Therefore, we have
\begin{eqnarray}
\pi^{\infty}_w &=& \frac{(1 - \delta)\pi^{0}_w + \delta P_t(\hat{q}_t)}{1 - \delta\left[P_w(1,1) - P_t(\hat{q}_t)\right]} \nonumber \\
&=& \frac{(1 - \delta)\pi^{0}_w + \delta \left\{ (1 - \alpha_t) + \alpha_t [(1 - 2 \epsilon)\hat{q}_t + \epsilon]^N\right\}}{1 - \delta (1 - \beta_w\alpha_w \epsilon) + \delta \left\{ (1 - \alpha_t) + \alpha_t [(1 - 2 \epsilon)\hat{q}_t + \epsilon]^N\right\}} \nonumber \\
&\ge & \frac{(1 - \delta)\pi^{0}_w + \delta (1 - \alpha_t)}{1 - \delta + \delta \beta_w \alpha_w \epsilon + \delta(1 - \alpha_t)}. \nonumber
\end{eqnarray}
The last inequality holds since $[(1 - 2 \epsilon)\hat{q}_t + \epsilon]^N \ge 0$ and $\pi^{\infty}_w$ is monotonically increasing as the value of $[(1 - 2 \epsilon)\hat{q}_t + \epsilon]^N$ increases.

Next, we show that the stationary distribution $\pi^{\infty}_w$ will be reached. From (\ref{stateDistUpdate}) and (\ref{stationaryStateCond1}), we have
\begin{equation}
\pi^{n + 1}_w - \pi^{\infty}_w = \delta\left[P_w(1,1) - P_t(\hat{q}_t)\right](\pi^{n}_w - \pi^{\infty}_w). \nonumber
\end{equation}
Since $|\delta\left[P_w(1,1) - P_t(\hat{q}_t)\right]| < 1$, we have
\begin{equation}
\lim\limits_{n \rightarrow \infty}(\pi^{n}_w - \pi^{\infty}_w) = 0 \Rightarrow \lim\limits_{n \rightarrow \infty} \pi^{n}_w = \pi^{\infty}_w. \nonumber
\end{equation}
\end{IEEEproof}

From Proposition 4, we can see the lower bound of $\pi^{\infty}_w$ increases as $\pi^{0}_w$ increases. Since the larger $\pi^{\infty}_w$ means higher efficiency, requesters should choose $\pi^{0}_w = 1$ for optimal performance. Therefore, we have
\begin{equation}
\pi^{\infty}_w \ge 1 - \frac{\delta \beta_w \alpha_w \epsilon}{1 - \delta + \delta(1 - \alpha_t) + \delta \beta_w \alpha_w \epsilon}.
\end{equation}

When $\beta_w = 0$, i.e., only the reward consensus is employed at the working state, or in the ideal case of $\epsilon = 0$, we can conclude that $\pi^{\infty}_w = 1$. This implies that every newly entered worker will first work at the working state, choose to produce solutions with the highest quality as their best responses and keep on working in the working state until they leave the system. As a result, all workers will stay at the working state and are available to solve posted tasks.

On the other hand, when $\beta_w > 0$ and $\epsilon > 0$, although all workers will start with the working state and choose to produce solutions with quality $1$, a portion of them will be put into the training state due to validation mistakes of requesters. However, since the probability of error is usually very small, i.e., $\epsilon \ll 1$, we can still expect $\pi^{\infty}_w$ to be very close to $1$, which implies that the majority of workers will be at the working state.

\section{Simulation Results}
In this section, we conduct numerical simulations to examine properties of our proposed mechanism $\mathcal{M}_t$ and to compare its performance with that of the basic mechanisms $\mathcal{M}_c$ and $\mathcal{M}_a$. Throughout the simulations, we assume the following cost function for workers
\begin{equation}
c(q) = \frac{(q + \lambda)^2}{(\lambda + 1)^2},
\end{equation}
where $\lambda > 0$ is a parameter that controls the degree of sensitivity of a worker's cost to his action. In particular, the smaller $\lambda$ is, the more sensitive a worker's cost will be with respect to his actions. In addition, the cost of choosing the highest quality $1$ is normalized to be 1, i.e, $c(1)=1$. From the definition of $c(q)$, we also have $c(0) = \frac{\lambda^2}{(\lambda + 1)^2}$ and $c^{\prime}(1) = \frac{2}{(\lambda + 1)}$. Moreover, we set $d = 10$, $\delta = 0.9$ and $\epsilon = 0.01$ throughout the simulations.

\begin{figure}
\centering
\includegraphics[height=2.6in, width=3.5in]{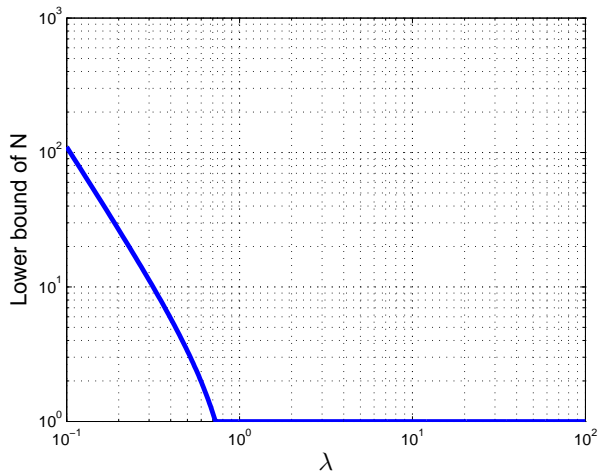}
\centering \caption{The lower bound of N for the existence of desirable symmetric Nash equilibria when $\beta_w = 0$.}
\end{figure}

In the first simulation, we evaluate the sufficient condition for the existence of desirable symmetric Nash equilibria in (\ref{lowerBound1}) under different settings. Such a sufficient condition is expressed in the form of a lower bound on the number of required training tasks, which depends on the worker's cost function as well as working state parameters $\beta_w$, $\alpha_w$ and $r$. We set $r = 1$, which matches the cost of producing solutions with quality $1$. Moreover, since $N \ge 1$, when the derived lower bound of N is less than $1$, we set it to be $1$ manually.

\begin{figure}
\centering
\includegraphics[height=2.6in, width=3.5in]{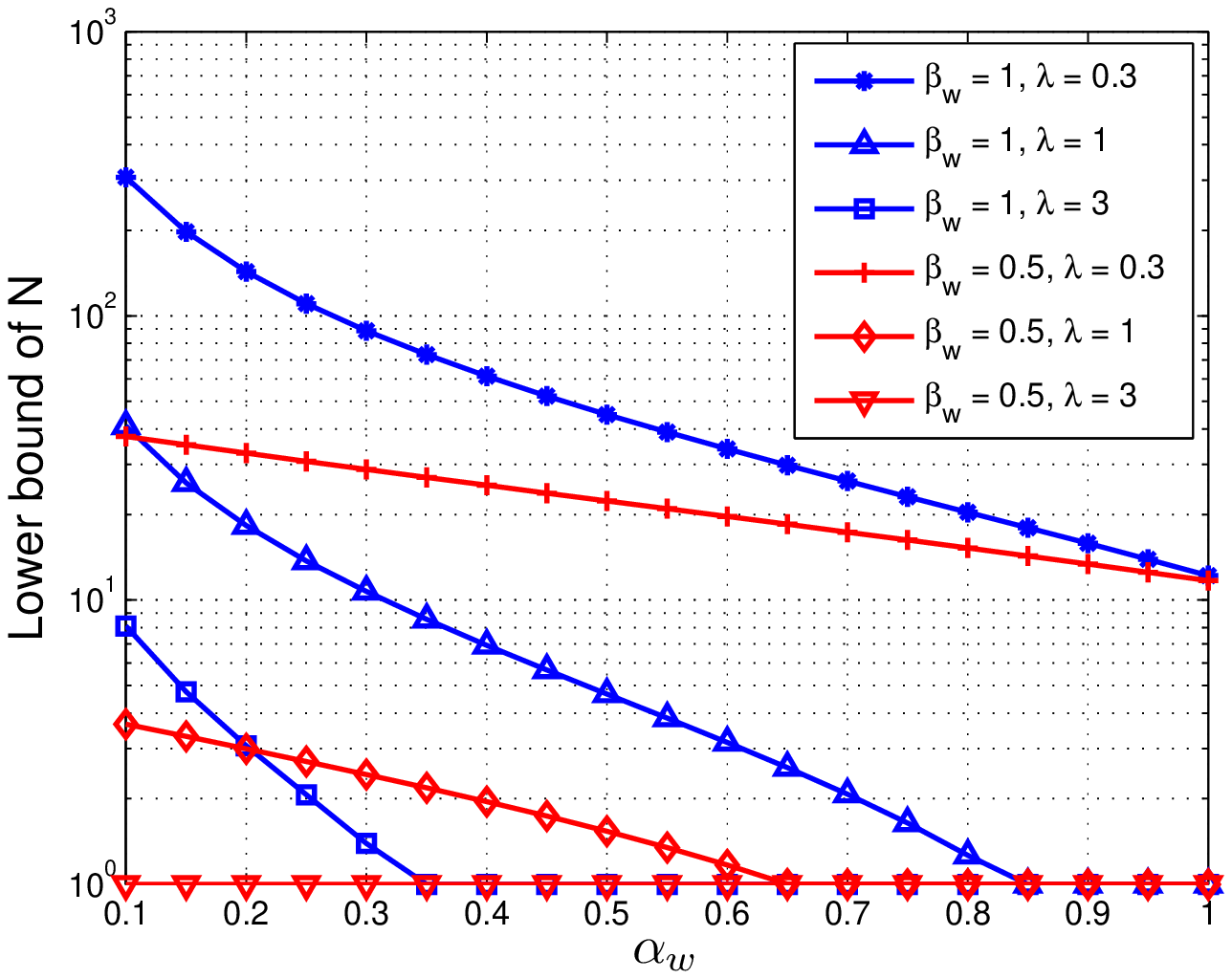}
\centering \caption{The lower bound of N for the existence of desirable symmetric Nash equilibria when $\beta_w \ne 0$.}
\end{figure}

We show in Fig. 2 the lower bound of $N$ versus $\lambda$ when $\beta_w = 0$, i.e., only the reward consensus mechanism is used in the working state.
Since workers are more cost-sensitive in producing high quality solutions with a smaller $\lambda$, it becomes more difficult to make $q = 1$ as their best responses. As a result, we need to set relatively large $N$s to achieve the desirable symmetric Nash equilibrium for small $\lambda$s as shown in Fig. 2. On the other hand, when $\lambda$ is large enough, the lower bound in (\ref{lowerBound1}) will no longer be an active constraint since any $N \ge 1$ can achieve our design objective.

We then study the more general cases where both the reward consensus mechanism and the reward accuracy mechanism are adopted in the working state. We show in Fig. 3 the lower bound of $N$ versus $\alpha_w$ under different values of $\beta_w$ and $\lambda$. Similarly, we can see that smaller $\lambda$ leads to a larger lower bound of $N$. Moreover, the lower bound of $N$ also increases as $\alpha_w$ decreases. This is due to the fact that it becomes more difficult to enforce workers to submit high quality solutions if we evaluate the submitted solutions less frequently. Since $\beta_w$ represents the ratio of tasks that will be evaluated using the reward accuracy mechanism, the smaller $\beta_w$ is, the less dependent of the lower bound of $N$ will be on the sampling probability $\alpha_w$.

\begin{figure}
\centering
\includegraphics[height=2.8in, width=3.5in]{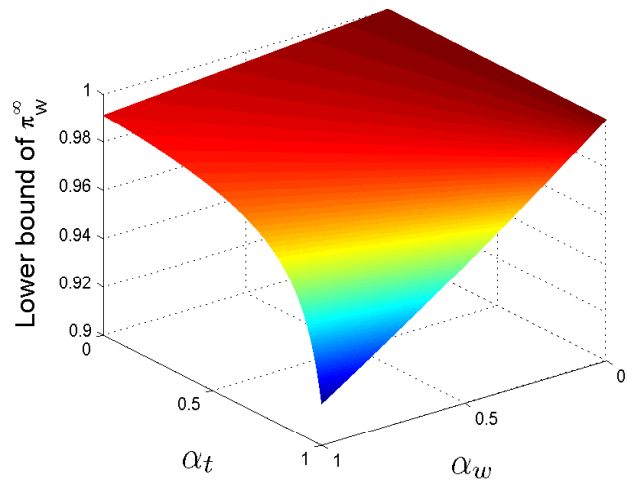}
\centering \caption{The lower bound of $\pi^{\infty}_w$ when $\beta_w = 1$.}
\end{figure}

In the second simulation, we evaluate numerically the lower bound of the stationary probability of a worker being at the working state, i.e., $\pi^{\infty}_w$ under different settings. We consider $\beta_w = 1$ in our simulations as $\pi^{\infty}_w = 1$ when $\beta_w = 0$. In addition,we set $\pi^{0}_w = 1$, i.e., every newly entered worker will be placed at the working state. In Fig. 4, we show the lower bound of $\pi^{\infty}_w$ under different values of $\alpha_w$ and $\alpha_t$. We can see that the lower bound of $\pi^{\infty}_w$ decreases as $\alpha_w$ and $\alpha_t$ increases. More importantly, $\pi^{\infty}_w$ will be above $0.9$ even in the worst case, which indicates that our proposed mechanism can guarantee the majority of workers being at the working state.

\begin{figure}[t]
\centering
  \subfigure[]
{\resizebox{5.2cm}{!}{\includegraphics[height=3.5cm,width=4cm]{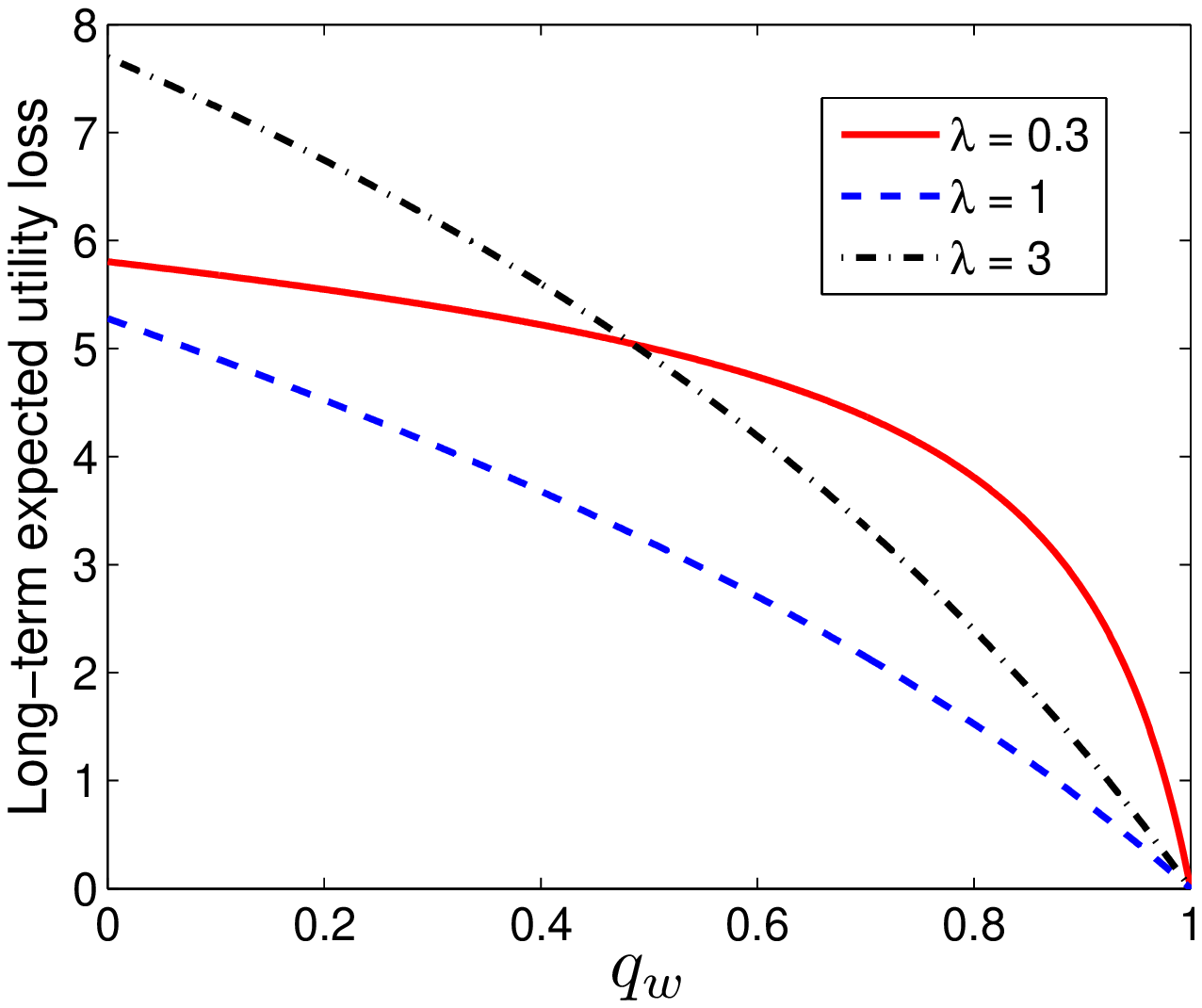}}} \hspace{0.1cm}
\subfigure[]
{\resizebox{5.2cm}{!}{\includegraphics[height=3.5cm,width=4cm]{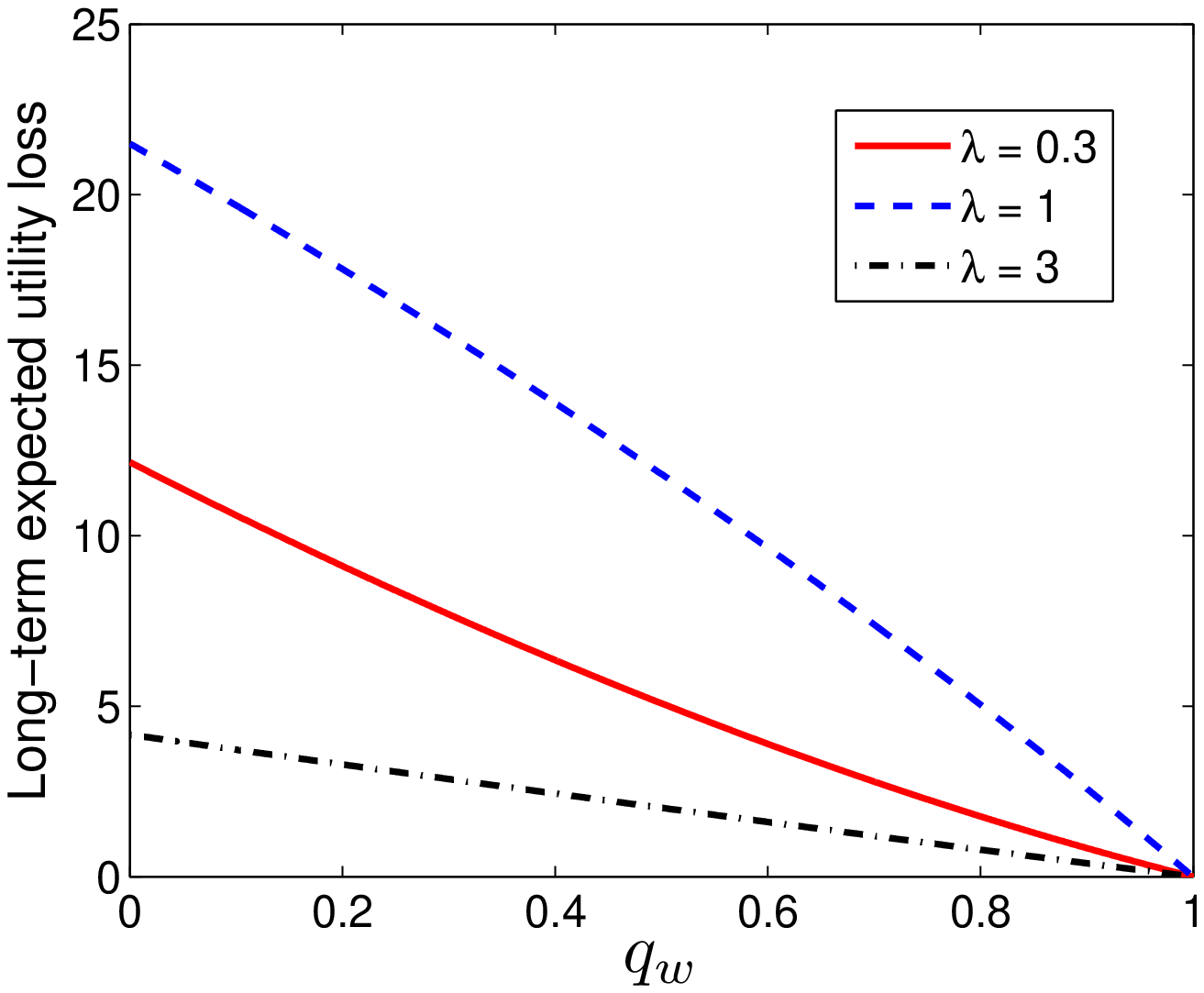}}} \hspace{0.1cm}
\subfigure[]
{\resizebox{5.2cm}{!}{\includegraphics[height=3.5cm,width=4cm]{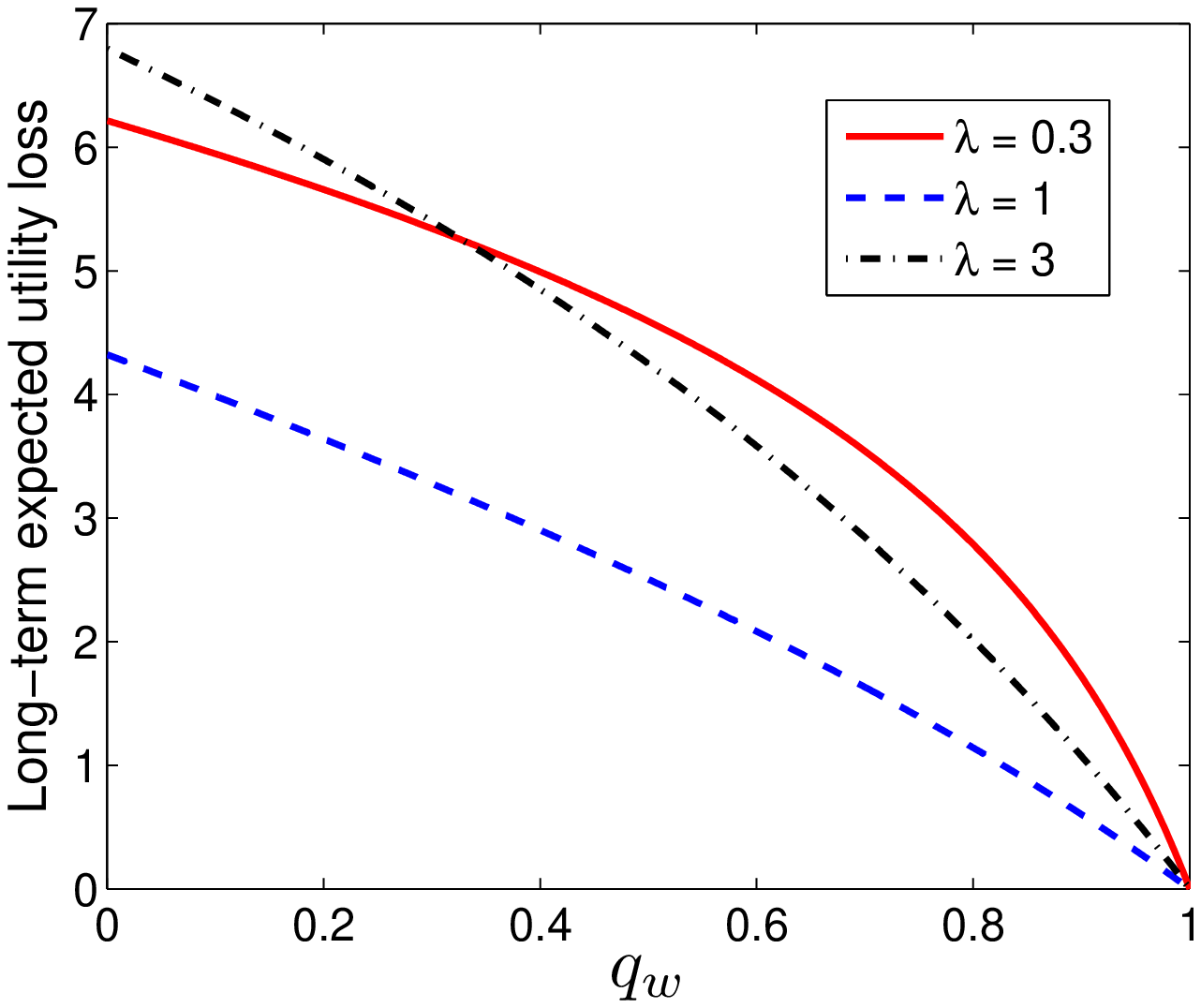}}}
\centering \caption{The long-term expected utility loss of a worker who deviates to action pair $(q_w, \hat{q}_t)$: (a) $\beta_w = 0$; (b) $\beta_w = 1$, $\alpha_w = 0.1$; (c) $\beta_w = 1$, $\alpha_w = 0.9$. }
\end{figure}

Next, we verify Theorem 1 through numerical simulations. In particular, we assume all workers adopt the equilibrium action pair $(1, \hat{q}_t)$ except one worker under consideration who may deviate to $(q_w, \hat{q}_t)$. We set $r = 1$ and choose $N$ to be the smallest integer that satisfies the sufficient condition of the existence of desirable symmetric Nash equilibria in (\ref{lowerBound1}). We set $\alpha_t$ according to (\ref{alpha_t2}) with $\gamma = 1$, i.e.,
\begin{equation}
\alpha_t = \min\left\{\frac{ \{3r(1 - \beta_w) + \beta_w\left[(1-\alpha_w\epsilon)r + \alpha_w d \right]\} }{ (1 - \epsilon^N) \{3r(1 - \beta_w) + \beta_w\left[(1-\alpha_w\epsilon)r + \alpha_w d \right]\} + \beta_w\alpha_w\epsilon N d}, 1\right\}.  \nonumber
\end{equation}
Moreover, the equilibrium action at the training state, $\hat{q}_t$, is obtained by solving (\ref{optEqn1}) and (\ref{optEqn2}) using the well-known value iteration algorithm [\ref{mdp}]. We show in Fig 5 the long-term expected utility loss of the worker under consideration at the working state, i.e., $U^{w}_{\mathcal{M}_t}(1, 1, \hat{q}_t) - U^{w}_{\mathcal{M}_t}(1, q_w, \hat{q}_t)$. From the simulation results, we can see that under all simulated settings, choosing $q_w = 1$ will always lead to the highest long-term expected utility, i.e., zero long-term expected utility loss. Therefore, as a rational decision maker, this worker will have no incentive to deviate from the action $(1, \hat{q}_t)$, which demonstrates that $(1, \hat{q}_t)$ is indeed sustained as an equilibrium.

\begin{figure}
\centering
\includegraphics[height=2.8in, width=3.5in]{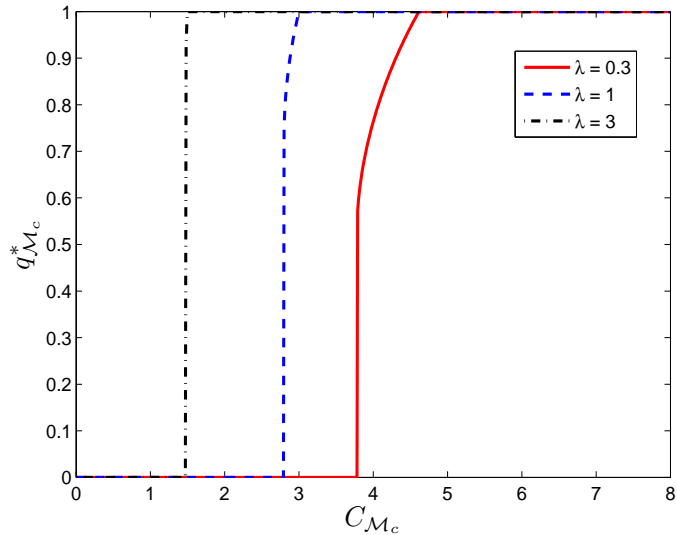}
\centering \caption{The equilibrium action versus the mechanism cost in $\mathcal{M}_c$.}
\end{figure}

Finally, we compare the performance of our proposed mechanism $\mathcal{M}_t$ with that of the two basic mechanisms $\mathcal{M}_c$ and $\mathcal{M}_a$. Since $\mathcal{M}_t$ is capable of incentivizing workers to submit solutions of quality $1$ with an arbitrarily low cost, it suffices to show the quality of solutions achieved by $\mathcal{M}_c$ and $\mathcal{M}_a$ under different mechanism costs. In particular, for $\mathcal{M}_c$, we assume that a task is given to $3$ workers. Therefore, for a given mechanism cost $C_{\mathcal{M}_c}$, the reward to each worker is $r = C_{\mathcal{M}_c}/3$. According to our analysis in Section \ref{MDMechanism}, the equilibrium action $q^*_{\mathcal{M}_c}$ in $\mathcal{M}_c$ can be calculated as $q^*_{\mathcal{M}_c} = \max\{\min\{q,1\},0\}$, where $q$ is the solution to the following equaiton
\begin{equation}
r[2q - q^{2}] = c^{\prime}(q). \nonumber
\end{equation}
In our simulations, when there are multiple equilibria, we pick the one with higher quality. On the other hand, if there exits no equilibrim, we set $q^*_{\mathcal{M}_c} = 0$. We show curves of the equilibrium action $q^*_{\mathcal{M}_c}$ in Fig. 6. From the simulation results, we can see that $\mathcal{M}_c$ can only achieve the highest quality $1$ when the mechanism cost $C_{\mathcal{M}_c}$ is larger than a certain threshold. Moreover, such a threshold increases as $\lambda$ increases, i.e., as workers are more cost sensitive in producing high quality solutions.

\begin{figure}[t]
\centering
\subfigure[]
{\resizebox{8cm}{!}{\includegraphics[height=3in, width=3.6in]{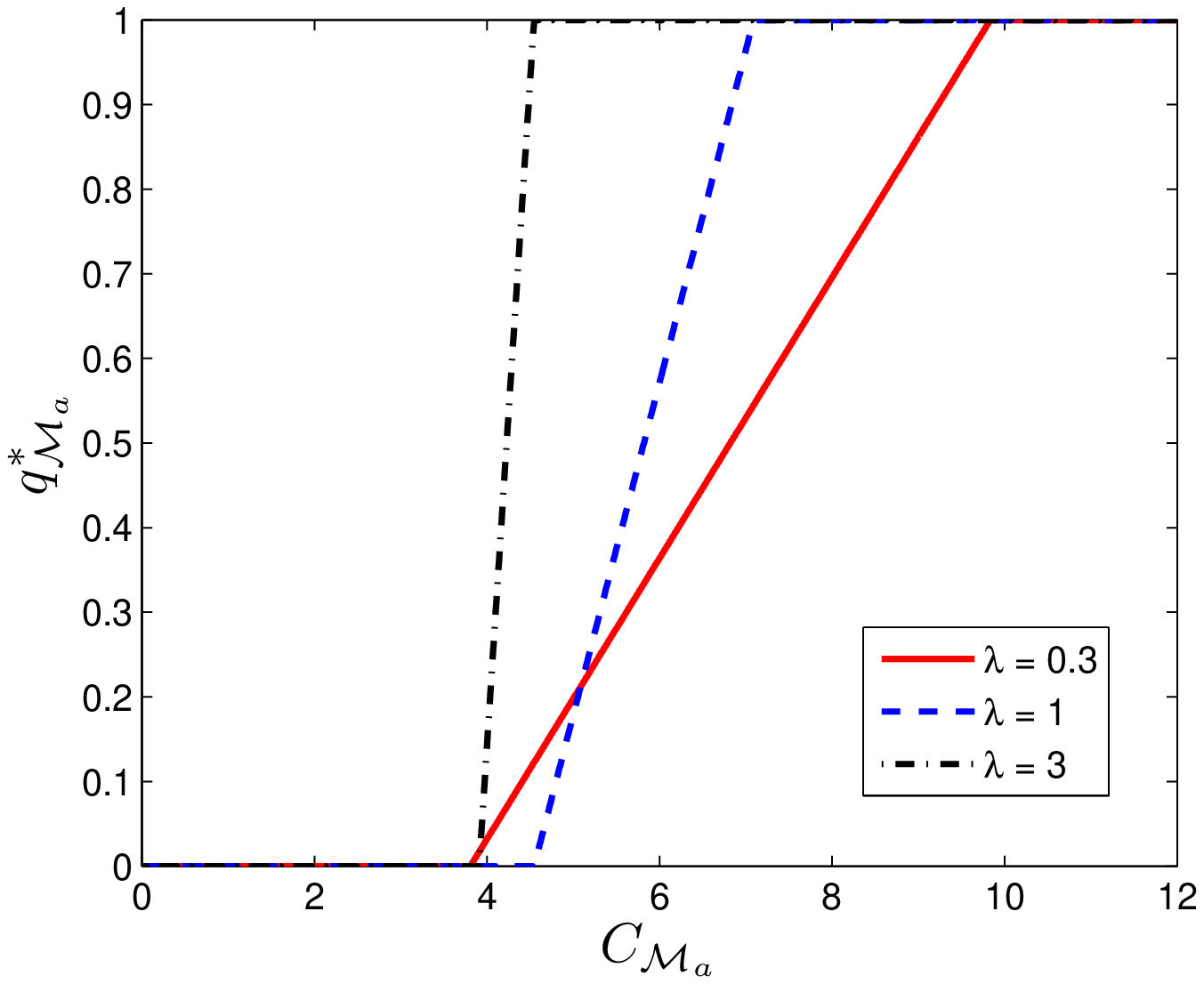}}} \hspace{0.2cm}
\subfigure[]
{\resizebox{8cm}{!}{\includegraphics[height=3in, width=3.6in]{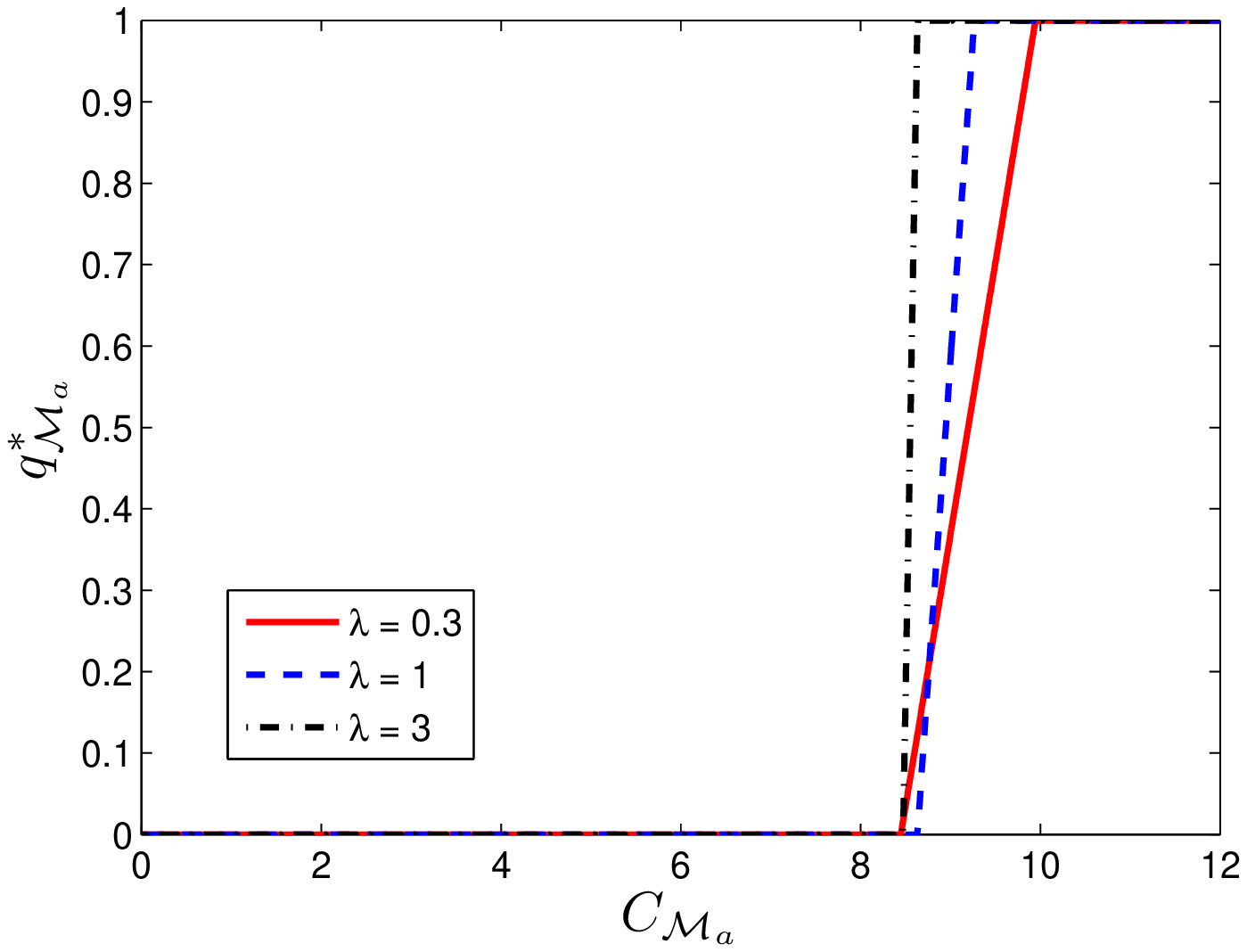}}}
\centering \caption{The optimal action versus the mechanism cost in $\mathcal{M}_a$: (a) $\alpha_a = 0.2$; (b) $\alpha_a = 0.8$. }
\end{figure}

For $\mathcal{M}_a$, we study two cases where $\alpha_a = 0.2$ and $\alpha_a = 0.8$, respectively. Then, given a mechanism cost $C_{\mathcal{M}_a}$, we set $r$ such that
\begin{equation}
C_{\mathcal{M}_a} = (1 - \alpha_a\epsilon)r + \alpha_a d. \nonumber
\end{equation}
Under $\mathcal{M}_a$, workers will respond by choosing their optimal action $q^*_{\mathcal{M}_a}$ as
\begin{equation}
q^*_{\mathcal{M}_a} = \arg \max\limits_{q\in[0,1]} u_{\mathcal{M}_a}(q).   \nonumber
\end{equation}
We show the optimal action $q^*_{\mathcal{M}_a}$ versus the mechanism cost $C_{\mathcal{M}_a}$ for $\mathcal{M}_a$ in Fig. 7. Similarly, we can see that requesters are unable to obtain high quality solutions with low $C_{\mathcal{M}_a}$.

\section{Experimental Verifications}
Beyond its theoretical guarantees, we further conduct a set of behavioral experiments to test our proposed incentive mechanism in practice. We evaluate the performance of participants on a set of simple computational tasks under different incentive mechanisms. We mainly focused on the reward accuracy mechanism in the experiment. We found that, through the use of quality-aware worker training, our proposed mechanism can greatly improve the performance of a basic reward accuracy mechanism with a low sampling probability to a level that is comparable to the performance of the basic reward accuracy mechanism with the highest sampling probability. We describe the experiment in detail below followed by analysis and discussions of the results.

\subsection{Description of The Experiment}
The task we used was calculating the sum of two randomly generated double-digit numbers. To make sure all tasks are of the same difficulty level, we further make the sum of unit digits to be less than $10$, i.e., there is no carry from the unit digits. The advantage of such a computational task is that: (a) it is straightforward for participants to understand the rule, (b) each task has a unique correct solution, (c) the task can be solved correctly with reasonable amount of effort, and (d) it is easy for us to generate a large number of independent tasks.

In our experiment, participants solve the human computation tasks in exchange for some virtual points, e.g., 10 points for each accepted solution. Their goal is to maximize the accumulated points earned during the experiment. Tasks are assigned to each participant in three sets. Each set has a time limit of 3 minutes and participants can try as many tasks as possible within the time limit. Such a time limit helps participants to quantify their costs of solving a task with various qualities using time. Different sets employ different incentive mechanisms. In particular, Set I employs the basic reward accuracy mechanism $\mathcal{M}_a$ with the highest sampling probability $\alpha_a = 1$. The basic reward accuracy mechanism $\mathcal{M}_a$ with a much lower sampling probability $\alpha_r = 0.3$ is employed in Set II. We use our proposed mechanism $\mathcal{M}_t$ in Set III, which introduces quality-aware worker training to the same basic reward accuracy mechanism as used in Set II with training state parameters set as $\alpha_r = 0$ and $N = 15$. Since correct solution can be obtained for all tasks, we are able to determine the correctness of each solution without error. That is, we have $\epsilon = 0$ in all cases.

We created a software tool to conduct the experiment. As no interaction among participants is involved, our experiment was conducted on an individual basis. Before the experiment, each participant was given a brief introduction to experiment rules as well as a demonstration of the software tool. There was also an exit survey followed each trial of the experiment, which asked participants about their strategies.

\subsection{Experimental Results}
We have successfully collected results from $41$ participants, most of whom are engineering graduate students. The number of collected submissions per set varies significantly from 30 to 180, depending on both the strategy and skills of different participants. From the requester's perspective, the accuracy of each participant represents the quality of submitted solutions and therefore is a good indicator to the effectiveness of incentive mechanisms. We show the histogram of accuracy for all three sets in Fig. 8.

\begin{figure}[t]
\centering
  \subfigure[]
{\resizebox{5.2cm}{!}{\includegraphics[height=3cm,width=3.5cm]{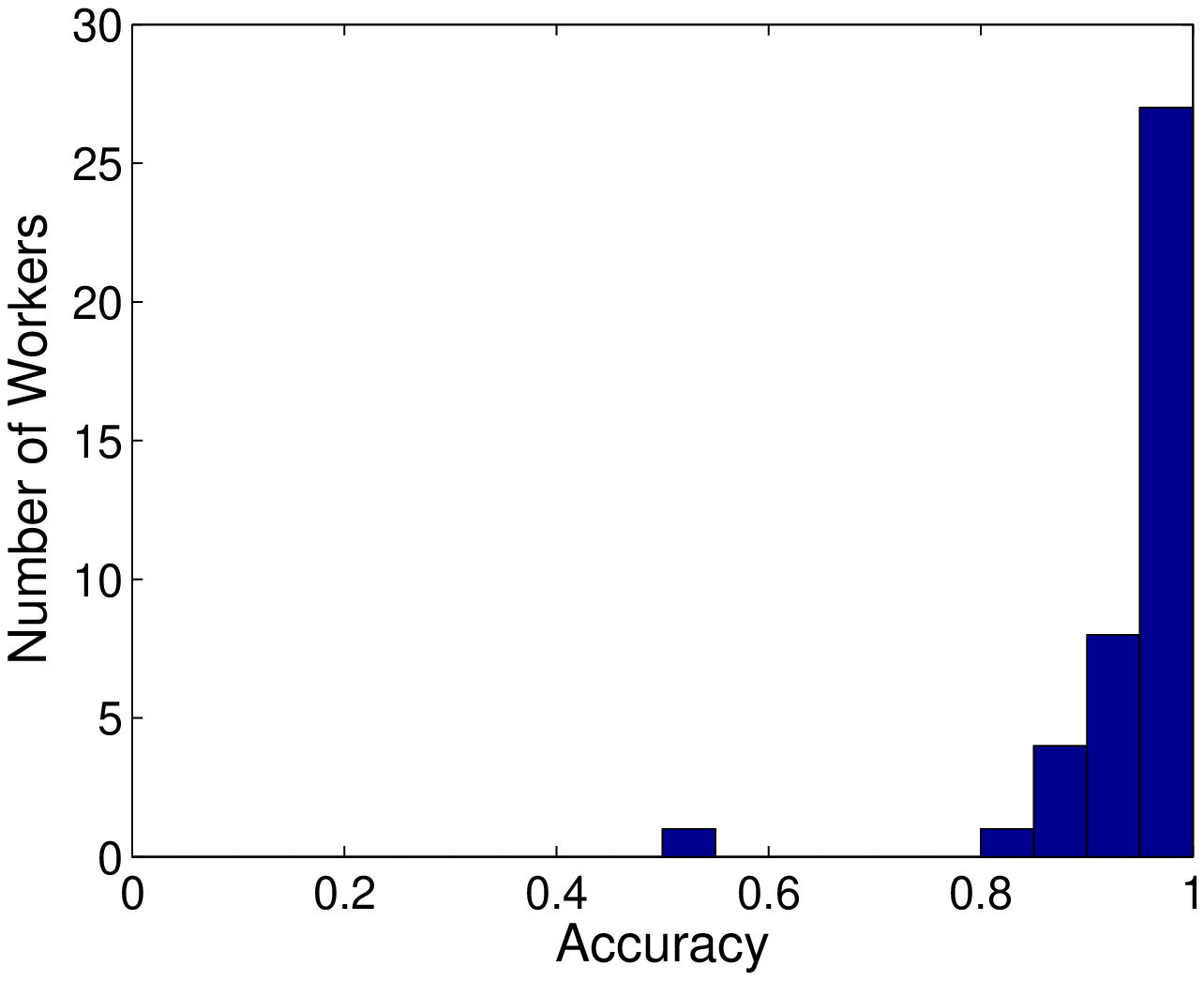}}} \hspace{0.1cm}
\subfigure[]
{\resizebox{5.2cm}{!}{\includegraphics[height=3cm,width=3.5cm]{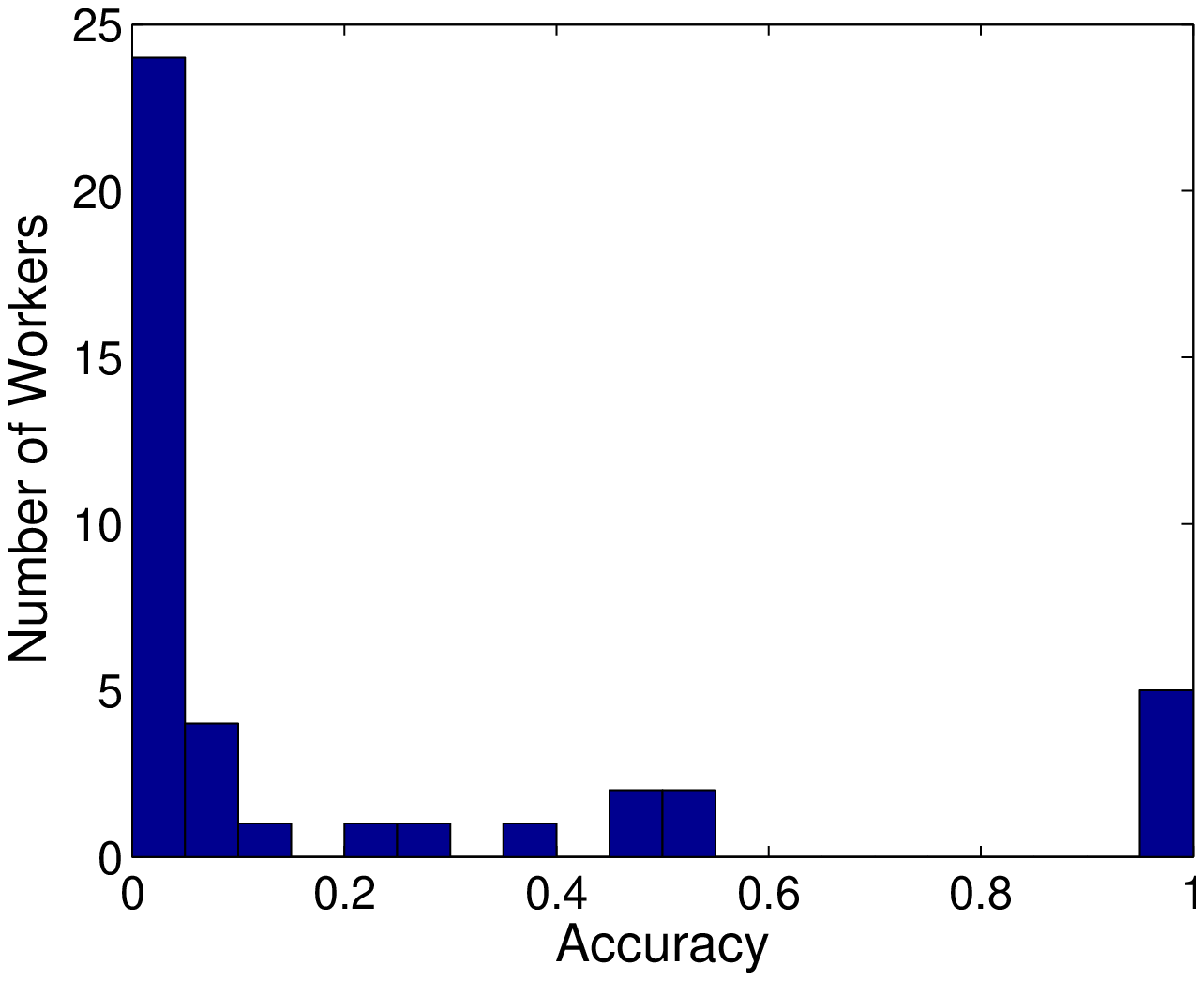}}} \hspace{0.1cm}
\subfigure[]
{\resizebox{5.2cm}{!}{\includegraphics[height=3cm,width=3.5cm]{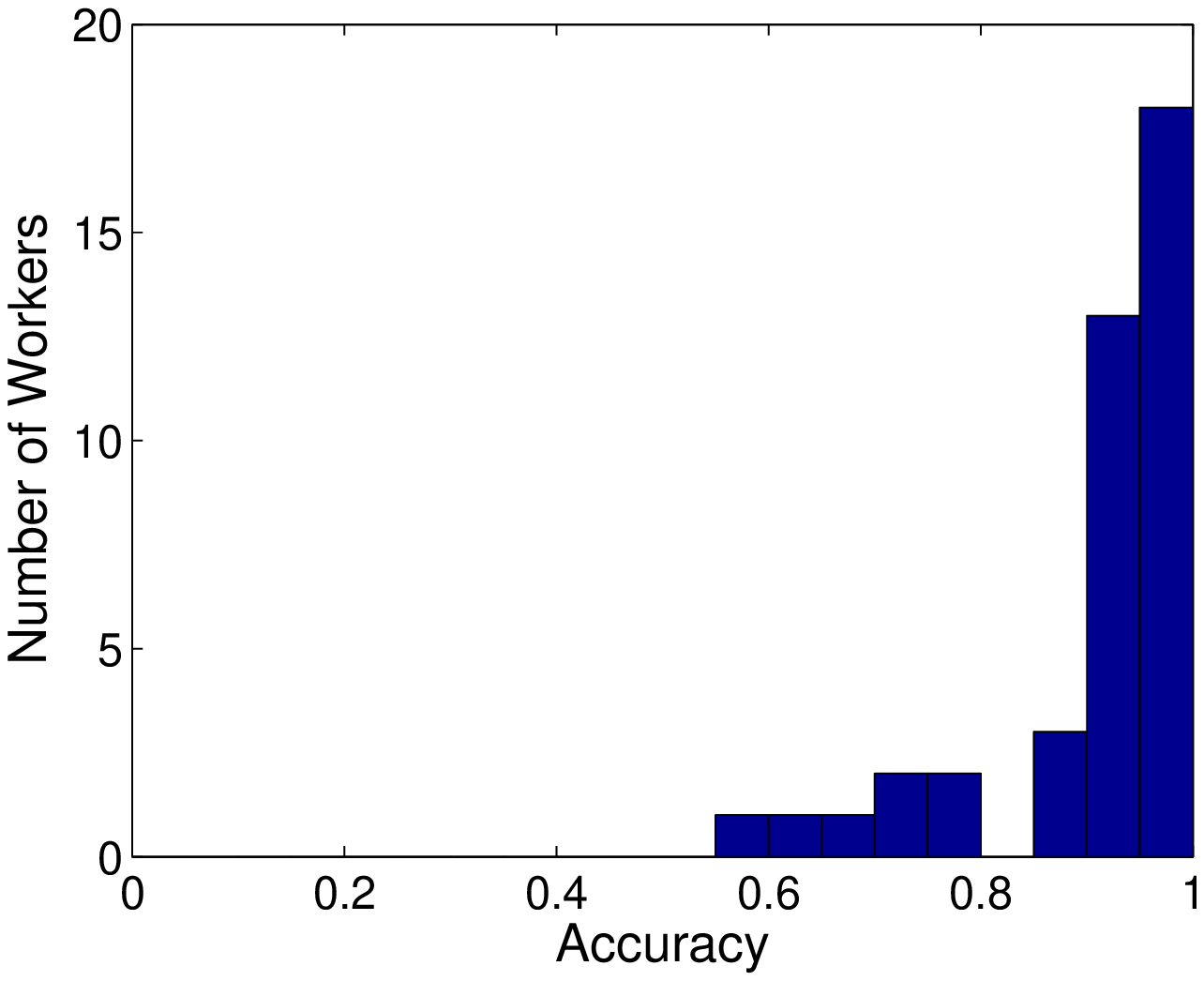}}}
\centering \caption{Histogram of accuracy: (a) Set I; (b) Set II; (c) Set III. }
\end{figure}

For Set I, as the highest sampling probability, i.e., $\alpha_a = 1$, was adopted, most participants responded positively by submitting solutions with very high qualities. There is only one participant who had relatively low accuracy compared with others in that he was playing the strategy of ``avoiding difficult tasks'' according to our exit survey. A much lower sampling probability of $0.3$ was used for Set II. In this case, it becomes profitable to increase the number of submissions by submitting lower quality solutions, as most errors will simply not be detected. This explains why the majority of participants had very low accuracies for Set II. Noteworthily, a few workers, 5 out 41, still exhibited very high accuracies in Set II. Our exit survey suggests that their behaviors are influenced by a sense of ``work ethics'', which prevents them to play strategically to exploit the mechanism vulnerability. Similar observations have also been reported in [\ref{Gao2012}] and [\ref{Paolacci2010}]. In Set III, as the introduction of training tasks make it more costly to submit wrong solutions, participants need to reevaluate their strategies to achieve a good tradeoff between accuracy and the number of submitted tasks. From Fig. 8, we can see that the accuracy of participants in Set III has a very similar distribution as that in Set I.

We now analyze our experimental results qualitatively. Let $\Gamma_I$, $\Gamma_{II}$ and $\Gamma_{III}$ represent the accuracy of Set I, Set II and Set III, respectively. Our results show that $\Gamma_{III} - \Gamma_{II}$ follows a distribution with median significantly greater than $0.6$ by the Wilcoxon signed rank test with significance level of $\rho < 5\%$. On the other hand, the median of the distribution of $\Gamma_{I} - \Gamma_{III}$ is not significantly greater than $0.01$ by the Wilcoxon signed rank test with $\rho \ge 10\%$. The unbiased estimate of the variance of $\Gamma_I$, $\Gamma_{II}$ and $\Gamma_{III}$ are $0.0060$, $0.1091$ and $0.0107$, respectively. Moreover, according to the Levene's test with significance level of $5\%$, the variance of $\Gamma_{III}$ is not significantly different from that of $\Gamma_{I}$ while it is indeed significantly different from that of $\Gamma_{II}$. To summarize, through the use of quality-aware worker training, our proposed mechanism can greatly improve the effectiveness of the basic reward accuracy mechanism with a low sampling probability to a level that is comparable to the one that has the highest sampling probability.

\section{Conclusions}
In this paper, we study cost-effective mechanisms for microtask crowdsourcing. In particular, we first consider two basic mechanisms widely adopted in existing microtask crowdsourcing applications and show that, to obtain high quality solutions, their mechanism costs must be higher than some lower bounds. Such lower bounds are beyond the control of requesters and may be high enough to negate the advantage of microtask crowdsourcing. Then, we propose a cost-effective mechanism based on quality-aware worker training. We prove theoretically that, given an arbitrarily low cost, our proposed mechanism can be designed to sustain a desirable equilibrium where workers choose to produce solutions with the highest quality at the working state and a worker will be at the working state with a large probability. Beyond its theoretical guarantees, we further conduct a set of human behavior experiments to demonstrate the effectiveness of our proposed mechanism.

\end{document}